\documentclass[superscriptaddress,reprint, amsmath,amssymb,aps,prl,nolongbibliography]{revtex4-2}
\usepackage[utf8]{inputenc}
\usepackage[T1]{fontenc}
\usepackage{graphicx}
\usepackage{dcolumn}
\usepackage{bm}
\usepackage{color,soul}
\usepackage{hyperref}
\usepackage{listings}
\usepackage{bbold}
\hypersetup{colorlinks=true,linkcolor=blue,citecolor=blue}
\newcommand{\be}{\begin{equation}}
\newcommand{\ee}{\end{equation}}
\newcommand{\bea}{\begin{eqnarray}}
\newcommand{\eea}{\end{eqnarray}}
\begin{document}

\title{Moiré excitons in biased twisted bilayer graphene under pressure}

\author{V. G. M. Duarte}
\email{vgmduarte@gmail.com}
\affiliation{Departamento de Física, Instituto Tecnológico de Aeronáutica, 12228-900, São José dos Campos, SP, Brazil}

\author{D. R. da Costa}
\email{diego_rabelo@fisica.ufc.br}
\affiliation{Departamento de Física, Universidade Federal do Ceará, 60455-900, Fortaleza, CE, Brazil}

\author{N. M. R. Peres}
\email{peres@fisica.uminho.pt}
\affiliation{Centro de Física das Universidades do Minho e do Porto (CF-UM-UP) e Departamento de Física, Universidade do Minho, P-4710-057 Braga, Portugal}
\affiliation{International Iberian Nanotechnology Laboratory (INL), Av Mestre José Veiga, 4715-330 Braga, Portugal}
\affiliation{POLIMA—Center for Polariton-driven Light-Matter Interactions, University of Southern Denmark, Campusvej 55, DK-5230 Odense M, Denmark}

\author{L. K. Teles}
\email{lara.teles@gp.ita.br}
\affiliation{Departamento de Física, Instituto Tecnológico de Aeronáutica, 12228-900, São José dos Campos, SP, Brazil}

\author{A. J. Chaves}
\email{andrejck@ita.br}
\affiliation{Departamento de Física, Instituto Tecnológico de Aeronáutica, 12228-900, São José dos Campos, SP, Brazil}

\date{\today}

\begin{abstract}
Using the tight-binding model, we report a gap opening in the energy spectrum of the twisted bilayer graphene under the application of pressure, that can be further amplified by the presence of a perpendicular bias voltage. The valley edges are located along the K-$\Gamma$ path of the superlattice Brillouin Zone, with the bandgap reaching values up to $200$ meV in the single-particle picture. Employing the formalism of the semiconductor Bloch equations, we observe an enhancement of the bandgap due to the electron-electron interaction, with a renormalization of the bandgap of about $160$ meV. From the solution of the corresponding Bethe-Salpeter equation, we show that this system supports highly anisotropic bright excitons whose electrons and holes are strongly hybridized between the adjacent layers. 
\end{abstract}

\maketitle

\textit{Introduction.---}Moiré patterns naturally appear when overlaying crystals with different individual lattice parameters or even in homobilayer when they are slightly offset due to rotation \cite{he2021moire, rakib2022moire}, such as turbostratic graphite owed to orientation disorder \cite{Johnson_1969}. Several extraordinary phenomena have been studied in different twisted materials, such as superconductivity and flat bands in what is called magic angle in twisted bilayer graphene (TBG) \cite{Cao2018, lisi2021observation, utama2021visualization}, Mott-like insulating states in half-filling TBG \cite{Cao2018x}, room temperature ferroelectricity in twisted bilayer MoS$_2$ \cite{Weston2022}, an alternation between ferromagnetic to antiferromagnetic domains in twisted bilayer CrI$_3$ \cite{Xiaodong2021}, and Hubbard physics in twisted bilayer WSe$_2$ \cite{Xu2022}.

For vertical stacked two-dimensional (2D) semiconductor materials with a twist, e.g., rotated transition metal dichalcogenides' (TMDs) bilayers, the periodic long-range interaction, known as moiré potential \cite{MacDonald2017}, results in a modulation of the band edge energies \cite{Shih2017}. In addition, tremendous interest in exciton physics bloomed with the advent of 2D materials due to their high binding energies \cite{mak2013tightly}, stemming from reduced screening and lower dimensionality \cite{Thygesen_2015}. In this context, recent experiments have demonstrated that such structural moiré patterns can trap long-lived and valley-polarized interlayer excitons, referred to as moiré excitons. \cite{tartakovskii2020excitons, seyler2019signatures, alexeev2019resonantly, jin2019observation, tran2019evidence} Those excitons have a wide range of possible applications, including the development of arrays of quantum emitters \cite{Wang2017} and excitonic devices \cite{ciarrocchi2022excitonic}.

Since 2006, the bandgap tunability of the AB-stacked bilayer graphene with a perpendicular electric field has been well-documented \cite{ohta2006controlling, CastroNeto2007, zhang2009direct}. This discovery triggered fundamental interest in exploring optical properties dominated by bound states. An additionally reported route to open a bandgap in biased Bernal stacked bilayer graphene is by nano-mechanical control, achieved, for instance, via interlayer distance decrease \cite{Guo2008}. The electrostatic control of electronic TBG bands and the possibility of bandgap tuning were theoretically explored \cite{talkington2023electric} for sublattice-exchange-dependent commensurate TBG with different interlayer shift vectors. 
Despite theoretical predictions \cite{park2010tunable, cheianov2012gapped, McEuen2017, Henriques2022} of tunable excitons in AB-stacked bilayer graphene as early as 2010, and experimental demonstrations \cite{McEuen2017} in 2017 showing the formation of excitons with large binding energies and distinct optical selection rules, excitons in TBG have to date not been neither predicted nor observed. 

Therefore, motivated by the experimental advances in the twist physics on the moiré excitons hosted in artificially engineered homobilayer \cite{vanderZande2014} and heterobilayer \cite{Tang2021} TMDs, that have revealed twist angle dependence in the excitonic properties and whose layer hybridization can be controlled by external electric field \cite{Tang2021}, in this Letter, we examine routes on the formation of excitons in TBG. To do that, using a tight-binding model within a single-particle picture, we first demonstrate gap opening in TBG under pressure and bias voltage for certain twist angles that generate commensurate unit cells, which suggests the possibility of exciton formation in bilayer graphene with a twist. Our analysis reveals that the single-particle bands are highly hybridized between different layers near the band edge. To appropriately describe the exciton formation, many-body interactions are added to the model, leading to the dielectric function calculation in the context of the Random Phase Approximation (RPA) \cite{bohm1951collective, pines1952collective, bohm1953collective}, adapting the Adler-Wiser formula \cite{Adler1962, Wiser1963} to account for the polarized screening effect between the two layers in 2D systems. Next, one evaluates the exchange self-energy that results in a bandgap energy correction, accounting for electron-electron Coulomb interaction, into the normalized optical band \cite{kira2011semiconductor, vanhala2020constrained}. As shall be discussed, our results clearly show a six-fold symmetric optical band with six nonequivalent band edges, where the lowest-energy exciton wavefunctions are localized.

\textit{Bandgap opening in single-particle TBG spectrum.---}We model commensurable TBGs as rigid and periodic 2D lattices, i.e. with restricted twist angles given by \cite{CastroNeto2012}: 
\begin{flalign}
\theta(p,q)=\arccos\left(\frac{3p^2+3pq+q^2/2}{3p^2+3pq+q^2}\right),
\end{flalign}
where $p,q$ are co-prime positive integers (see Sec.~S1 in Supplemental Material \cite{supplementary}). It was recently shown \cite{Lin2023} that the geometric relaxation of atoms can be safely neglected for twist angles above $1.8^\circ$, which theoretically ensures us to limit discussing large angle cases with rigid rotation features. In this context, the non-interacting tight-binding Hamiltonian for TBG in the presence of a perpendicularly applied electric field, taking solely the low-energy dominant $p_z$ orbitals, is represented in the momentum space, as follows (see Sec.~S2 in Supplemental Material \cite{supplementary})
\begin{equation}\label{eq:H0 subs b to a}
    \mathcal{H}_0 \hspace{-0.1cm} = \hspace{-0.1cm}\sum_{\mathbf{k}}^{\text{BZ}} \hspace{-0.05cm}\sum_{\ell\ell'} \sum_{\boldsymbol{\delta}_{\ell}\boldsymbol{\delta}_{\ell'}} \hspace{-0.05cm} \sum_{nn'} h_{\boldsymbol{\delta}_{\ell}\boldsymbol{\delta}_{\ell'}}\hspace{-0.025cm}(\mathbf{k}) u^*_{n\boldsymbol{\delta}_{\ell}}\hspace{-0.025cm}(\mathbf{k}) u_{n'\boldsymbol{\delta}_{\ell'}}\hspace{-0.025cm}(\mathbf{k}) a^\dagger_{n\mathbf{k}} a_{n'\mathbf{k}},
\end{equation}
where $u_{n\boldsymbol{\delta}_{\ell}}(\mathbf{k})$ is the Bloch wavefunctions that form an orthonormal basis, $\ell$ and $n$ are the layer- and band-indices, respectively, $\boldsymbol{\delta}_\ell$ is the basis vectors of the superlattice, $a_{n\mathbf{k}}^\dagger = \sum_{\boldsymbol{\delta}_{\ell}} u_{n\boldsymbol{\delta}_{\ell}}(\mathbf{k}) b_{\mathbf{k}\boldsymbol{\delta}_{\ell}}^\dagger$ is the Bloch operator, $\sum_{\mathbf{k}}^{\text{BZ}}$ is a sum over wavevectors restricted to the first Brillouin Zone (BZ), and the matrix element $h_{\boldsymbol{\delta}_{\ell}\boldsymbol{\delta}_{\ell'}}(\mathbf{k})$ is defined as
\begin{equation}\label{eq:matrix element}
    h_{\boldsymbol{\delta}_{\ell}\boldsymbol{\delta}_{\ell'}}\hspace{-0.025cm}(\mathbf{k}) \hspace{-0.05cm}=\hspace{-0.05cm} 
    \dfrac{V}{2} s_{\ell}\delta_{\ell\ell'}\delta_{\boldsymbol{\delta}_{\ell} \boldsymbol{\delta}_{\ell'}}
    \hspace{-0.025cm}+ \hspace{-0.025cm}\sum_{\mathbf{R}} e^{\mathrm{i}\mathbf{k}\cdot(\mathbf{R}+\boldsymbol{\delta}_{\ell}-\boldsymbol{\delta}_{\ell'})} t(\mathbf{R}+\boldsymbol{\delta}_{\ell}-\boldsymbol{\delta}_{\ell'}),
\end{equation}
with $s_{\ell}=\delta_{\ell 1}-\delta_{\ell 2}$, which incorporates the bias voltage contribution $V$ and the transfer integral $t(\mathbf{r})$. In terms of the Slater-Koster form \cite{SlaterKoster, Moon2013}, the transfer integrals, which describe the hopping energies between orbitals of two atoms, are modeled as exponential functions given by 
\begin{align}\label{eq:transfer integral}
    t(\mathbf{r})
    & = 
    V_{pp\sigma}(r)\left(\dfrac{\mathbf{r}\cdot\mathbf{e_z}}{r}\right)^2 
    + V_{pp\pi}(r) \left[1-\left(\dfrac{\mathbf{r}\cdot\mathbf{e_z}}{r}\right)^2\right],
\end{align}
where the pure transfer integrals $V_{pp\pi}(r)$ and $V_{pp\sigma}(r)$ correspond to intralayer ($\pi$--like) and interlayer ($\sigma$--like) C--C interactions (see Sec.~S2 of the Supplemental Material \cite{supplementary}).

To obtain the single-particle energies $E_n(\mathbf{k})$ and Bloch wavefunctions $u_{n\boldsymbol{\delta}_\ell}(\mathbf{k})$, we numerically construct the tight-binding Hamiltonian $\mathcal{H}_0$ in the reciprocal space of the commensurable TBG superlattice, whose dimension is equal to the number of atoms in the unit cell, and then one exactly diagonalizes $\mathcal{H}_0$ for each point $\mathbf{k}$. Here, the intralayer hoppings are restricted to nearest neighbors. The interlayer hoppings are included up to a cutoff $d_c$, ensured by the fact that C---C interactions decay exponentially with the distance, implying that one can safely truncate the infinite sum running over the lattice vectors $\sum_{\mathbf{R}}$ \cite{mirzakhani2020}. Under pressure, the interlayer distance decreases, resulting in an enhancement in the interlayer hopping energies. Here, the pressure effect on rigid TBG is mimicked by simply varying the interlayer atomic distance equally away from the equilibrium position $d_0 = 3.35$ \AA. No changes due to pressure are considered on the intralayer hoppings, i.e., we will neglect the Poisson effect.

\begin{figure}[t]
\centering
\includegraphics[width=1.01\columnwidth]{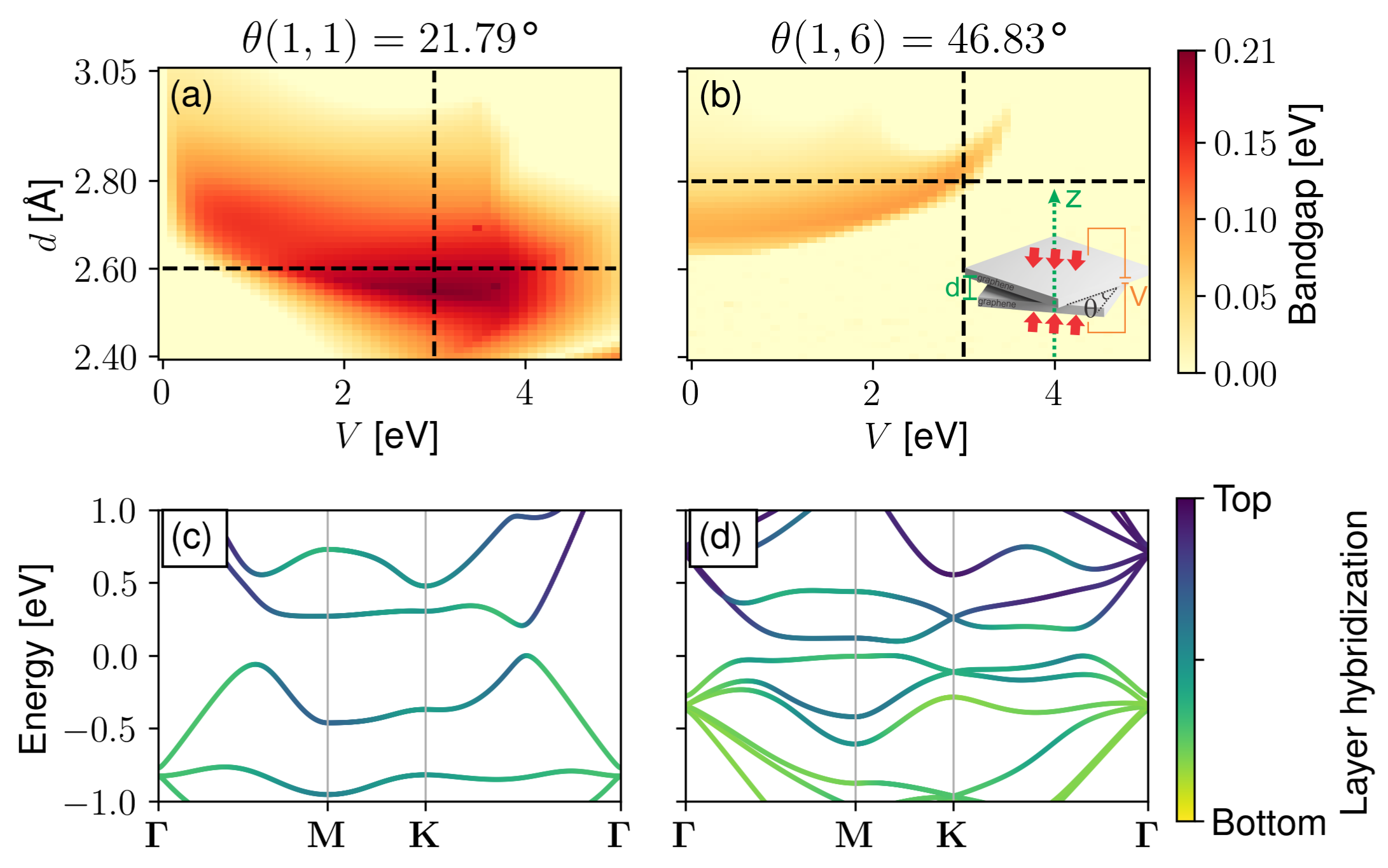}
\caption{(Color online) Electric field and pressure-induced bandgap opening in TBG with a twist angle (a) $\theta\left(1,1\right) = 21.79^\circ$ and (b) $\theta\left(1,6\right) = 46.83^\circ$. The corresponding electronic band structures along the path $\mathbf{\Gamma}-\mathbf{M}-\mathbf{K}-\mathbf{\Gamma}$ for the highest gap-opening values for a fixed bias potential of $V=3$ eV are shown in (c) with $E_g \approx 0.18$ eV and (d) with $E_g \approx 0.068$ eV, respectively, as marked by the crossing points of the dashed lines in panels (a) with $d=2.6$ \AA\ and (b) with $d=2.8$ \AA. The color map indicates the layer hybridization associated with the bottom-top wavefunction composition. The inset in (b) illustrates the TBG under pressure with an interlayer distance $d$ and bias potential $V$.}
\label{Fig1}
\end{figure}

Figures~\ref{Fig1}(a) and \ref{Fig1}(b) show color plots of the single-particle bandgap dependence on the interlayer distance $d$ and the applied electric field magnitude $V$ for a TBG with twist angle (a) $\theta\left(1,1\right) = 21.79^\circ$ and (b) $\theta\left(1,6\right) = 46.83^\circ$. Results for different twist angles, corresponding to TBG structures that generate commensurable unit cells of less than $500$ atoms, are given in Sec.~S3 of the Supplemental Material \cite{supplementary}. Our findings demonstrate, in certain cases, bandgaps up to $0.2$ eV, when one combines electric bias and pressure, although just one of them is enough to generate a gap opening in some cases, as shown in Fig.~S3 of the Supplemental Material \cite{supplementary}. Therefore, it turns TBG into a narrow gap semiconductor and, consequently, robust enough to support excitons. The electronic band structures and their layer composition, associated with the spatial localization of the electrons on the individual layers projected in each band, for a fixed bias potential $V = 3$ eV are depicted in Figs.~\ref{Fig1}(c) for an interlayer distance of $d=2.6$~\AA\ and twist angle of $\theta=21.79^\circ$ and \ref{Fig1}(d) for  $d=2.8$~\AA\  and $\theta=46.83^\circ$, revealing strong layer hybridization around the bands' edges in both cases. Thus, we will show that gapped TBG hosts layered-hybridized moiré excitons. Besides pressure, we could also consider the stretching of TBG layers, not included here, which would increase the interlayer over intralayer hopping ratios, favoring the appearance of a bandgap. Recently, it has been demonstrated \cite{talkington2023electric} that the sliding of one graphene layer over the other could open a bandgap.

\textit{Dieletric screening, excitons, and optical response for TBG.---} To properly describe electron-hole bound states, the screened interaction should be incorporated in the formalism \cite{Latini2015}. For that, we consider the static dielectric function within the RPA and use the Adler-Wiser formula \cite{Adler1962, Wiser1963} adapted here for 2D system (see Sec.~S3 in Supplemental Material \cite{supplementary}), resulting in
\begin{align}\label{eq:die}
    &\epsilon^{\ell\ell'}_{\mathbf{G}\mathbf{G}'}(\mathbf{q}) = \delta_{\ell\ell'}\delta_{\mathbf{G}\mathbf{G}'} + \dfrac{e^2}{\mathcal{S}} \sum_{\ell''} X^{\ell\ell''}(|\mathbf{q}+\mathbf{G}|) \nonumber \\ 
    & \times \sum_{n_c n_v} \sum_{\mathbf{k}}^{\text{BZ}} \left[\dfrac{\left(M^{n_c n_v}_{\ell''}(\mathbf{k},\mathbf{q},\mathbf{G})\right)^* M^{n_c n_v}_{\ell'}(\mathbf{k},\mathbf{q},\mathbf{G}')}{E_{n_c \mathbf{k}} - E_{n_v \mathbf{k}+\mathbf{q}}} \right. \nonumber \\ 
    & \left. + \dfrac{\left(M^{n_v n_c}_{\ell''}(\mathbf{k},\mathbf{q},\mathbf{G})\right)^* M^{n_v n_c}_{\ell'}(\mathbf{k},\mathbf{q},\mathbf{G}')}{E_{n_c \mathbf{k}+\mathbf{q}} - E_{n_v \mathbf{k}}} \right],
\end{align}
where $\mathbf{G}$ is the reciprocal lattice vector of the superlattice, $n_c$ ($n_v$) is the conduction (valence) band-index, $\mathcal{S}$ is the surface area of the system, and the overlap term is defined as
\begin{equation}
M^{nn'}_{\ell}(\mathbf{k},\mathbf{q},\mathbf{G}) = \sum_{\boldsymbol{\delta}_{\ell}} u^*_{n\boldsymbol{\delta}_{\ell}}(\mathbf{k}) e^{\mathrm{i}\mathbf{G}\cdot\boldsymbol{\delta}_{\ell}} u_{n'\boldsymbol{\delta}_{\ell}}(\mathbf{k}+\mathbf{q}),
\end{equation}
being $u_{n\boldsymbol{\delta}_{\ell}}(\mathbf{k})$ the Bloch wavefunction component of the $p_z$ localized in the $\boldsymbol{\delta}_\ell$ site of the supercell. The term $X^{\ell\ell'}$ is associated with the bare electron-electron interaction, obtained by solving the Poisson equation. It is written as
\begin{equation}
X^{\ell \ell^\prime}(k)
= \dfrac{\delta_{\ell \ell^\prime} + e^{-kd}(1 - \delta_{\ell\ell^\prime})}{2\epsilon_0 k},
\end{equation}
where $\ell=\ell^\prime$ and $\ell\neq\ell^\prime$ contributions correspond to the intralayer and interlayer interactions, respectively. In the long-wavelength approximation, the dielectric function \eqref{eq:die} becomes $\epsilon^{\ell\ell^\prime}_{\mathbf{0}\mathbf{0}} \approx \delta_{\ell\ell'}+r^{\ell\ell^\prime}_0q$, that corresponds to the Rytova-Keldysh potential with a screening length $r_0^{\ell\ell^\prime}$ \cite{Cudazzo2011}. Thus, by numerically calculating the dielectric function \eqref{eq:die} and taking the first-order expansion term (see Sec.~S3 and Fig.~S4 in Supplemental Material \cite{supplementary}), one gets $r_0\approx 208.2$ \AA, and $r_0\approx 340.6$ \AA, for the intralayer contribution in the twist angles of $\theta(1,1) = 21.79^\circ$ and $\theta(1,1) = 46.83^\circ$, respectively, being them higher than the $r_0$ value for monolayer TMDs \cite{Pedersen2016}. The obtained $r_0$ values obey the trend reported in Ref.~\cite{Zhenyu_2015} that the smaller the bandgap, the higher $r_0$ value. The effective interaction, taking into account the screening of p$_z$ orbitals of the twisted structure in each layer, is given by
\be
\psi_{\mathbf{G}\mathbf{G}^\prime}^{\ell_1\ell_2}(\mathbf{q}) =
\sum_{\ell^\prime}
{\epsilon^{-1}}^{\ell_2\ell^\prime}_{\mathbf{G}\mathbf{G}^\prime}(\mathbf{q})
X^{\ell_1\ell^\prime}(|\mathbf{q}+\mathbf{G}'|),
\ee 
that shows a different screening for interlayer and intralayer potentials (see Sec.~S5 in Supplemental Material \cite{supplementary}). 

\begin{figure}[t!]
\centering
\includegraphics[width=1\columnwidth]{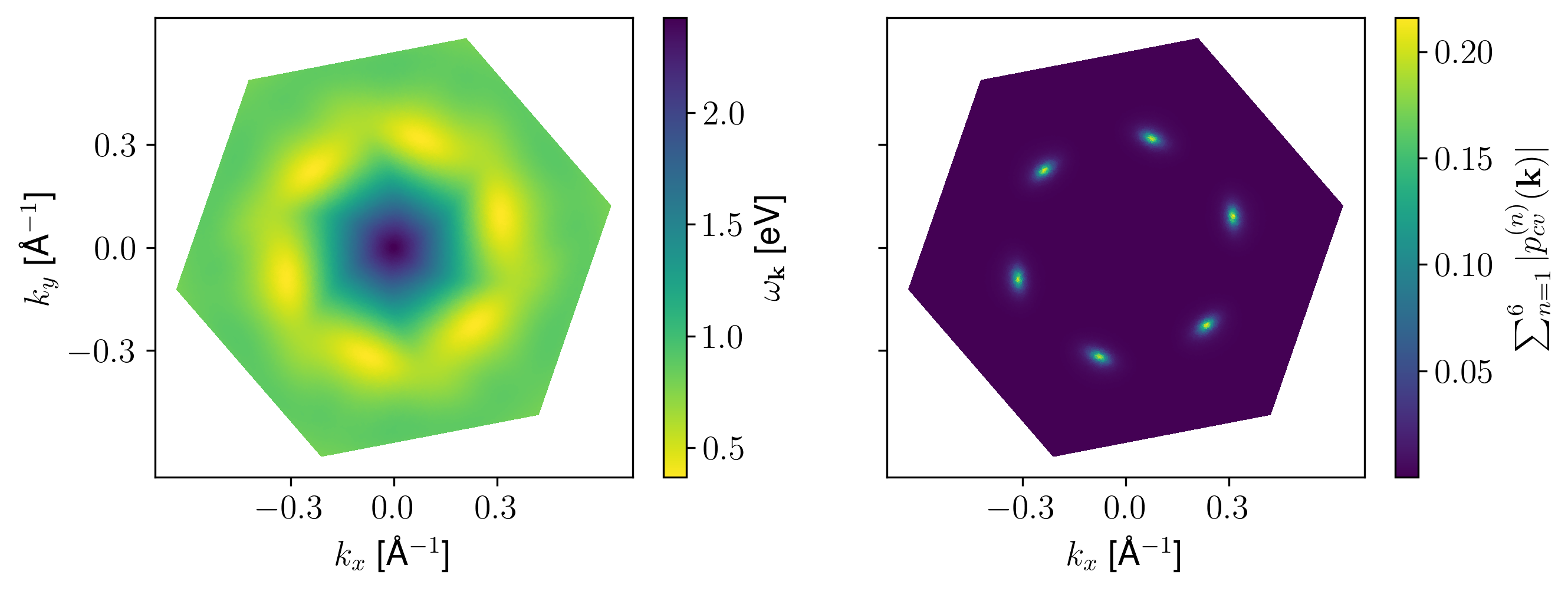}
\caption{(Color online) (a) Renormalized optical band structure, exhibiting a six-fold minimum in the first Brillouin zone of the supercell, and (b) a merging plot of the six first excitonic wavefunctions in momentum space, located exactly in the optical band edges, for a TBG with $\theta(1,1)=21.79^\circ$, $V=3$ eV, and $d=2.6$ \AA.}
\label{Fig2}
\end{figure}

To study the exciton properties using the tight-binding formalism, we employ the Semiconductor Bloch Equations (SBE) \cite{kira2011semiconductor}, as explained in detail in Sec.~S5 in Supplemental Material \cite{supplementary}. The SBE is obtained by writing the Heisenberg equation of motion for the interband transition amplitude $p_\mathrm{cv}(\mathbf{k})=\langle \hat{a}^\dagger_{\mathrm{c},\mathbf{k}} \hat{a}_{\mathrm{v},\mathbf{k}} \rangle$, with $\hat{a}_{\mathrm{n},\mathbf{k}}$ being the annihilator operator for an electron at the band $n$ with wavenumber $\mathbf{k}$ for the incidence of an electromagnetic wave with frequency $\omega$ and amplitude $\boldsymbol{\mathcal{E}}$. We neglected the Auger process, considering only the highest valence (v) and lowest conduction bands (c), and used the rotating-wave approximation and equilibrium occupation numbers for $T=0$. In this case, the SBE read as
\begin{align}
\left(\hbar\omega \hspace{-0.075cm}-\hspace{-0.075cm}\hbar\tilde{\omega}_{\mathbf{k}} \hspace{-0.075cm}+ \hspace{-0.075cm}i\gamma\right)p_\mathrm{cv}(\mathbf{k})\hspace{-0.075cm}+\hspace{-0.15cm}\int \hspace{-0.15cm}\frac{d^2\mathbf{q}}{(2\pi)^2}K(\mathbf{k},\mathbf{q})p_\mathrm{cv}(\mathbf{q})  \hspace{-0.035cm}= \hspace{-0.025cm} \mathbf{d}_\mathrm{vc}(\mathbf{k})\cdot \boldsymbol{\mathcal{E}}, \label{eq:sbe}
\end{align}
where $\hbar\tilde{\omega}_{\mathbf{k}}$ is the renormalized transition energy with the inclusion of the exchange self-energy $\hbar\tilde{\omega}_{\mathbf{k}} = E_{c\mathbf{k}} - E_{v\mathbf{k}} + \Sigma_{\mathbf{k}}$, $\gamma$ is a phenomenological term for the relaxation transition rate, the integral is performed over the first Brillouin zone, $\mathbf{d}_\mathrm{vc}(\mathbf{k})$ is the dipole matrix element, and the Kernel $K(\mathbf{k},\mathbf{q})$ is
\begin{align}
K(\mathbf{k},\mathbf{q})= e^2 \sum_{\ell_1\ell_2} \sum_{\mathbf{G}\mathbf{G}^\prime} \psi_{\mathbf{G}\mathbf{G}^\prime}^{\ell_1\ell_2}(\mathbf{q} &-\mathbf{k}) \left(M_{\ell_1}^{cc}(\mathbf{k},\mathbf{q}-\mathbf{k},\mathbf{G}') \right)^*\nonumber\\
&\times M_{\ell_2}^{vv}(\mathbf{k},\mathbf{q}-\mathbf{k},\mathbf{G}),
\end{align}
with the exchange self-energy read as:
\begin{align}\label{eq:self-energy}
    \hspace{-0.2cm} \Sigma_{\mathbf{k}} & \hspace{-0.075cm}=\hspace{-0.075cm} \dfrac{e^2}{\mathcal{S}} \sum_{\ell_1\ell_2} \sum_{\mathbf{G}\mathbf{G}'} \sum_{\mathbf{q}}^{\text{BZ}} \psi^{\ell_1\ell_2}_{\mathbf{G}\mathbf{G}'}(\mathbf{q}) \left[ (M^{vv}_{\ell_1}(\mathbf{k},\mathbf{q},\mathbf{G}'))^* \right.\nonumber\\ & \hspace{-0.25cm} \times \hspace{-0.1cm} \left. M^{vv}_{\ell_2}(\mathbf{k},\mathbf{q},\mathbf{G}) \hspace{-0.075cm}-\hspace{-0.075cm} (M^{cv}_{\ell_1}(\mathbf{k},\mathbf{q},\mathbf{G}'))^* \hspace{-0.05cm} M^{vc}_{\ell_2}(\mathbf{k},\mathbf{q},\mathbf{G})
    \right].
\end{align}
For the calculation of exciton states, we solve the homogeneous version of Eq.~\eqref{eq:sbe}, setting $\boldsymbol{\mathcal{E}}=\mathbf{0}$ and $\hbar\omega=E_n$ to be the corresponding eigenvalue.

From the numerically calculated Bloch functions $u_{n\delta_\ell}(\mathbf{k})$ and using the inverse of the dielectric function \eqref{eq:die}, we obtain the exchange self-energy term and then the renormalized optical band $\hbar\omega_\mathbf{k}$. Such results are shown in Fig.~\ref{Fig2}(a) for $\theta(1,1)=21.79^\circ$, $V=3$ eV, and $2.6$ \AA ~and for $\theta(1,6) = 46.83^\circ$, $V=3$ eV and $d=2.8$ \AA ~in Fig.~S5(a) in Supplemental Material \cite{supplementary}. For $\theta(1,1)=21.79^\circ$, the renormalization of the bandgap was almost of $120$ meV, resulting in bandgaps up to $364$ meV; however, no pronounced qualitative changes were observed from the transition energy $\hbar\omega_\mathbf{k}$. From Fig.~\ref{Fig2}(a), one clearly notices a six-fold minimum in the $\hbar\tilde{\omega}_\mathbf{k}$ spectrum located on the band edges along the $K$--$\Gamma$ path in the first Brillouin zone of the supercell. In Fig.~\ref{Fig2}(b), we show the superposition of the six first excitonic wavefunctions in momentum space, each of them located exactly at one of the six minima of the optical band. They are also strongly anisotropic, spread along the direction orthogonal to the $K$--$\Gamma$ path. In Sec.~S5 of Supplemental Material \cite{supplementary}, we present the layer hybridization's calculation of the excitonic bound states, showing that both the electrons and holes are highly hybridized between the two layers.

\begin{figure}[t]
\centering
\includegraphics[width=1.01\columnwidth]{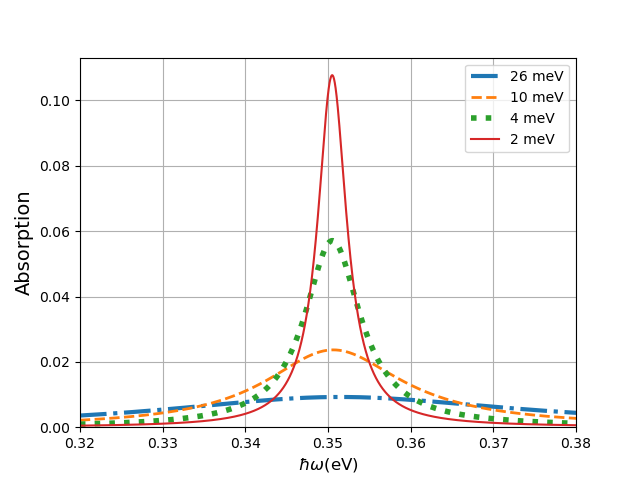}
\caption{(Color online) Light absorption for different values of the relaxation rate $\gamma$. The parameters of the TBG are $\theta(1,1)=21.79^\circ$, $V=3$ eV, and $d=2.6$ \AA.}
\label{Fig3}
\end{figure}

For bright excitons, reflectance and absorption measurements will show signatures of their presence  \cite{Li2014}. From the SBE \eqref{eq:sbe}, we can compute the optical conductivity and, therefore, the absorption \cite{Chaves_2017}. As we are interested in the frequencies near the exciton resonance, we use Elliot's formula for the optical conductivity (see Sec.~S5 in the Supplemental Material \cite{supplementary})
\begin{equation}
\sigma(\omega)=4i\omega\sigma_0 \sum_j \frac{{\tilde{\mathbf{d}}^*_j \otimes \tilde{\mathbf{d}}_j}}{\omega-\omega_n+i\gamma/\hbar},
\end{equation}
with $\otimes$ being the outer product, $\sigma_0=e^2/\hbar$, and  $\tilde{\mathbf{d}}_j$ the dimensionless exciton dipole moment
\begin{equation}
\tilde{\mathbf{d}}_j=\int\frac{d^2\mathbf{k}}{(2\pi)^2} \phi^*_j(\mathbf{k}) \mathbf{d}_{cv}(\mathbf{k}),
\end{equation}
where $\phi_j(\mathbf{k})$ are the normalized excitonic eigenmodes of the homogeneous version of Eq.~\eqref{eq:sbe}. From the optical conductivity $\sigma(\omega)$, we calculate the absorption for a suspended sample, where we include the contribution of the first six degenerated exciton states (see Fig.~\ref{Fig2}), and we show the results in Fig.~\ref{Fig3} for different values of the relaxation rate $\gamma$, that represents the effects of disorder and temperature. The absorption increases as the nonradiative decay rate $\gamma$ decreases, reaching values up to 10\% absorption for $\gamma=2$ meV; a value that is compatible with excitons in TMDs \cite{Epstein2020}. Thus, we show that optical measurements can probe the presence of excitons in this system.

\textit{Conclusion.---}In summary, we demonstrated that a bandgap can be induced in TBG under pressure for realistic conditions when a voltage bias is applied. In this gapped TBG case, we predicted the existence of highly anisotropic moiré excitons that strongly interact with light. Moreover, the methodology developed here for studying excitons in moiré systems can be applied to other 2D twisted materials. 

\section*{Acknowledgments} 

V.~G.~M.~D. acknowledges a Msc. scholarship from the Brazilian agency CAPES (Fundação Coordenação de Aperfeiçoamento de Pessoal de N\'ivel Superior). A.~J.~C., L.~K.~T, and D.~R.~C. were supported by CNPq (Conselho Nacional de Desenvolvimento Cient\'ifico e Tecnol\'ogico) Grant No.~400789/2019-0, 310422/2019-1, 308486/2015-3, 310422/2019-1, 315408/2021-9, 423423/2021-5, 408144/2022-0, and 313211/2021-3.
N.M.R.P. acknowledges support from the Independent Research Fund Denmark (grant no. 2032-00045B) and the Danish National Research Foundation (Project No. DNRF165).
A.~J.~C. acknowledges Funda\c{c}\~ao de Amparo \`a Pesquisa do Estado de S\~ao Paulo (FAPESP) under Grant No.~2022/08086-0, and is kindly grateful to INL (International Iberian Nanotechnology Laboratory) for the warm hospitality, where part of this work was carried out.

\bibliography{ref_new}

\end{document}


\title{Supplemental Material for ``{Moiré excitons in biased twisted bilayer graphene under pressure''}}

\author{V. G. M. Duarte}
\email{vgmduarte@gmail.com}
\affiliation{Departamento de Física, Instituto Tecnológico de Aeronáutica, 12228-900, São José dos Campos, SP, Brazil}

\author{D. R. da Costa}
\email{diego\_rabelo@fisica.ufc.br}
\affiliation{Departamento de Física, Universidade Federal do Ceará, 60455-900, Fortaleza, CE, Brazil}

\author{N. M. R. Peres}
\email{peres@fisica.uminho.pt}
\affiliation{Centro de Física das Universidades do Minho e do Porto (CF-UM-UP) e Departamento de Física, Universidade do Minho, P-4710-057 Braga, Portugal}
\affiliation{International Iberian Nanotechnology Laboratory (INL), Av Mestre José Veiga, 4715-330 Braga, Portugal}
\affiliation{POLIMA—Center for Polariton-driven Light-Matter Interactions, University of Southern Denmark, Campusvej 55, DK-5230 Odense M, Denmark}

\author{L. K. Teles}
\email{lara.teles@gp.ita.br}
\affiliation{Departamento de Física, Instituto Tecnológico de Aeronáutica, 12228-900, São José dos Campos, SP, Brazil}

\author{A. J. Chaves}
\email{AndreJ6@gmail.com}
\affiliation{Departamento de Física, Instituto Tecnológico de Aeronáutica, 12228-900, São José dos Campos, SP, Brazil}

\begin{abstract}
In this Supplemental Material file, we present in detail, in addition to those discussed in the main text and that support the physics presented there, the following interesting sections: \textbf{(S1) TBG crystallographic structure}, presenting the superlattice primitive vectors in real and reciprocal spaces and the commensurate condition when rotates two graphene layers with respect to each other, \textbf{(S2) TBG tight-binding model}, discussing the tight-binding Hamiltonian for TBG, the transfer integral associated with the hoppings energies, how to incorporate an external electric field on the model, and a detail derivation to write down the eigenvalue problem to be numerically diagonalized, \textbf{(S3) bandgap opening}, reporting a systematic study of the bandgap in TBG evaluated as a function of the bias potential $V$ and the interlayer distance $d$, where changes in the latter parameter mimic the application of vertical pressure to the system, \textbf{(S4) dielectric screening}, deriving analytically the static dielectric function of TBG in the context of the Random Phase Approximation, \textbf{(S5) many-body effects: exictons and optical response}, exhibiting the many-body formalism to derive the semiconductor Bloch equation and the renormalized optical band's equation, both in the context of the Random Phase Approximation, as well as additional results in relation to those of the main text for the optical band, binding energies, exciton wavefunctions, and its layer hybridization.
\end{abstract}

\maketitle

\section{TBG crystalographic structure}\label{sec:atomic structure}

\begin{figure}[t]
    \centering
    \includegraphics[width=1\columnwidth]{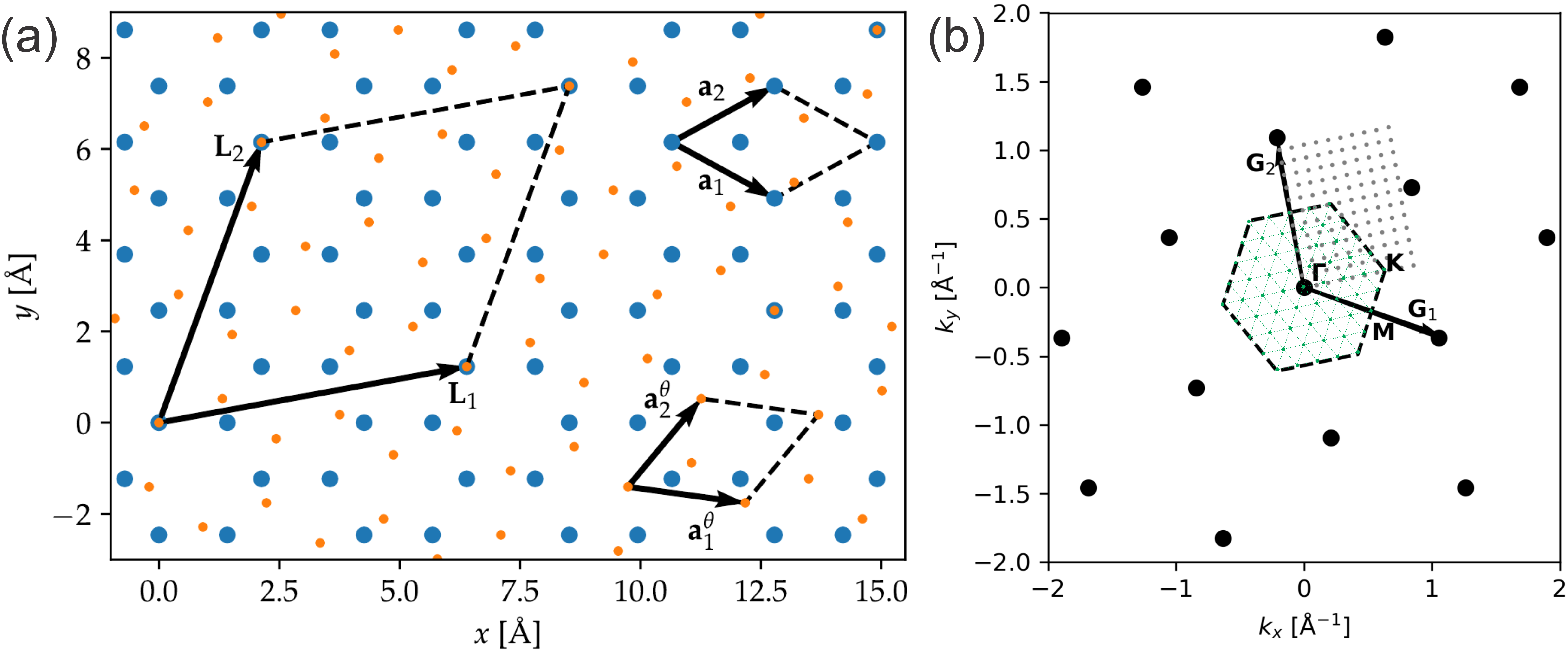}
    \caption{(Color online) (a) A top view schematic representation in real space of TBG for $\theta(1, 1) = 21.79^\circ$. Large blue and small orange symbols represent the atoms of the unrotated ($\ell=1$) and rotated ($\ell=2$) layers, respectively. The rhombus-like selected unit cell is highlighted using the primitive vectors $\mathbf{L}_1$, $\mathbf{L}_2$, and black dashed lines. The primitive vectors and unit cells of each monolayer are also shown. (b) TBG reciprocal lattice for $\theta(1, 1) = 21.79^\circ$. Big black dots denote the lattice points $\mathbf{G}=m\mathbf{G}_1 + n\mathbf{G}_2$ ($m,n\in\mathbb{Z}$), $\mathbf{G}_{1,2}$ are the primitive vectors in the momentum space. The small gray and green dots correspond to the discretized $k$ points of two equivalent first Brillouin zones with rectangular and hexagonal shapes, respectively. For illustrative purposes, unrealistically small values of sampling points were used [see $N_{x',y'}$ in Eq.~\eqref{eq:k-point sampling rectangular} for rectangular grid case]. Some high-symmetry points are also indicated: $\mathbf{\Gamma}$, $\mathbf{K}$, $\mathbf{M}$.}
    \label{fig:tbg_lattice}
\end{figure}

We model a TBG as two planar graphene layers, \textit{i.e.}, buckling effects are neglected. We choose a coordinate system in which the layers are located at $z=0$ and $z=d$, such that $d$ is the vertical distance between the layers. The unit vectors that form the standard basis of the 3D space are denoted as $\mathbf{e}_i$, with $i=\{x,y,z\}$. Each monolayer graphene lattice is composed of two triangular sublattices, A and B, and their crystalline orientations are rotated relative to each other by an angle $\theta$. Fig.~\ref{fig:tbg_lattice} shows the TBG lattice from the reference point of an observer looking from above, in the direction $-\mathbf{e}_z$.

The sublattice vectors for the unrotated layer ($\ell=1$ with sublattices $A_1$ and $B_1$) and the rotated layer ($\ell=2$ with sublattices $A_2$ and $B_2$) can be explicitly written, respectively, as
\begin{subequations} \label{eq:triangular lattices}
    \begin{align}
        \mathbf{r}_{A1} &= m\mathbf{a}_1 + n\mathbf{a}_2, \label{eq:sublattice A1}\\
        \mathbf{r}_{B1} &= \mathbf{r}_{A1} + (\mathbf{a}_1+\mathbf{a}_2)/3, \label{eq:sublattice B1}\\
        \mathbf{r}_{A2} &= \text{rot}(\theta)\mathbf{r}_{A1} + d\mathbf{e}_z, \label{eq:sublattice A2}\\
        \mathbf{r}_{B2} &= \text{rot}(\theta)\mathbf{r}_{B1} + d\mathbf{e}_z, \label{eq:sublattice B2}
    \end{align}
\end{subequations}
where $m,n \in \mathbb{Z}$, $\mathbf{a}_{\bfrac{1}{2}} = a(\sqrt{3}\mathbf{e}_x \mp \mathbf{e}_y)/2$ are the primitive vectors of $(\ell=1)$--graphene layer, as depicted in Fig.~\ref{fig:tbg_lattice}(a), $a=2.46$ Å is a lattice constant, and $\text{rot}(\theta)$ is the rotation matrix
\begin{equation}\label{eq:rotation matrix}
    \text{rot}(\theta) = \begin{bmatrix}
    \cos\theta & -\sin\theta & 0\\
    \sin\theta & \cos\theta & 0\\
    0 & 0 & 1
    \end{bmatrix}.
\end{equation}

The following analysis is restricted to TBG structures with well-defined superlattices and unit cells. In other words, the TBG superlattice must be periodic. This is guaranteed if some atomic positions of different layers match horizontally, which can be written mathematically as
\begin{equation}\label{eq:commensurate condition}
    m_1 \mathbf{a}_1 + n_1 \mathbf{a}_2 = m_2 \mathbf{a}_1 + n_2 \mathbf{a}_2,
\end{equation}
for some sets of integers $\{m_1,n_1,m_2,n_2\}$.
This Diophantine equation is known as the commensurate condition \cite{lopes2007,mele2010,shallcross2010,lopes2012}. Its solutions are given in terms of an arbitrary pair of co-prime positive integers $\{p,q\}$, such that the possible twist angles between graphene layers are
\begin{equation}\label{eq:commensuration condition}
    \theta(p,q) = \arccos\left(\dfrac{3p^2 + 3pq + q^2/2}{3p^2 + 3pq + q^2}\right).
\end{equation}
The primitive vectors of the resulting commensurated superlattices are given by
\begin{subequations}
    \begin{align}\label{eq:primitive vectors}
        \mathbf{L}_1 &=\begin{cases}
        \left(p+\dfrac{q}{3}\right)\mathbf{a}_1+\dfrac{q}{3}\mathbf{a}_2, &\text{if } $q$ \text{ is divisible by } 3,\\
        p\mathbf{a}_1+(p+q)\mathbf{a}_2, &\text{otherwise},
        \end{cases}\\
        \mathbf{L}_2 &=\text{rot}(60^\circ)\mathbf{L}_1.
    \end{align}
\end{subequations}

Here, we denote by $\mathbf{R}=m\mathbf{L}_1+n\mathbf{L}_2$ ($m,n\in\mathbb{Z}$) the superlattice vectors and by $\boldsymbol{\delta}_{\ell}$ the subset of basis vectors of the graphene layer $\ell=\{1,2\}$ that define the unit cell of the superlattice, as illustrated in Fig.~\ref{fig:tbg_lattice}(a). Mathematically, $\boldsymbol{\delta}_{\ell}$ is any lattice vector whose projection in the $xy-$plane can be written as
\begin{equation}
    x\mathbf{L}_1 + y\mathbf{L}_2, \, 0\leq x,y < 1.
\end{equation}

Any sublattice vector (\ref{eq:triangular lattices}) can be rewritten generically as $\mathbf{R}+\boldsymbol{\delta}_{\ell}$.
The compact notations $\sum_{\ell}=\sum_{\ell=1}^2$, $\sum_{\mathbf{R}}$ and $\sum_{\boldsymbol{\delta}_{\ell}}$ will be used to indicate sums over all layers, superlattice vectors, and basis vectors, respectively. The reciprocal lattice points $\mathbf{G}=m\mathbf{G}_1+n\mathbf{G}_2$ ($m,n\in\mathbb{Z}$), the reciprocal primitive vectors
\begin{subequations}
\begin{align}
    \mathbf{G}_1 &= 2\pi \dfrac{\mathbf{L}_2\times\mathbf{e}_z}{|\mathbf{L}_1\times\mathbf{L}_2|},\\
    \mathbf{G}_2 &= 2\pi \dfrac{\mathbf{e}_z\times\mathbf{L}_1}{|\mathbf{L}_1\times\mathbf{L}_2|},
\end{align}
\end{subequations}
and the first Brillouin zone with some of the high symmetry points (located at the hexagon center $\mathbf{\Gamma}=\mathbf{0}$, the vertices $\mathbf{K}$ and the edge midpoints $\mathbf{M}$) are depicted in Fig.~\ref{fig:tbg_lattice}(b).

\section{TBG tight-binding model}\label{sec.S2}

\subsection{TBG tight-binding Hamiltonian}\label{sec:TBG tight-binding Hamiltonian}

Our description is based on the tight-binding approximation for the energetically dominant TBG orbitals, $p_z$, in the vicinity of the energy levels on TBG band edges. The tight-binding Hamiltonian in the second quantization formalism can be written as
\begin{equation}\label{eq:tight-binding hamiltonian}
    \mathcal{H}_{\text{TB}} = \sum_{\ell\ell'} \sum_{\boldsymbol{\delta}_{\ell} \boldsymbol{\delta}_{\ell'}} \sum_{\mathbf{R}\mathbf{R}'} t(\mathbf{R}+\boldsymbol{\delta}_{\ell}-\mathbf{R}'-\boldsymbol{\delta}'_{\ell'}) c_{\mathbf{R}+\boldsymbol{\delta}_\ell}^\dagger c_{\mathbf{R}'+\boldsymbol{\delta}_{\ell'}},
\end{equation}
where $t(\mathbf{r})$ is the transfer integral \cite{SlaterKoster, Moon2013} and $c_{\mathbf{r}}^{\dagger}$ ($c_{\mathbf{r}}$) is the fermionic operator that creates (annihilates) a $p_z$ electron centered at $\mathbf{r}$. By assuring commensurability (see Sec.~\ref{sec:atomic structure}), the atomic structure of TBG is guaranteed to be periodic. To take advantage of this periodicity, we Fourier transform the operators $c^\dagger_{\mathbf{r}}$ using
\begin{equation}\label{eq:op fourier transform}
    c^\dagger_{\mathbf{R}+\boldsymbol{\delta}_{\ell}} = \dfrac{1}{\sqrt{N}} \sum_{\mathbf{k}}^{\text{BZ}} e^{\mathrm{i}\mathbf{k}\cdot(\mathbf{R}+\boldsymbol{\delta}_{\ell})} b_{\mathbf{k}\boldsymbol{\delta}_{\ell}}^\dagger,
\end{equation}
where $N$ is the number of unit cells of the material and $\sum_{\mathbf{k}}^{\text{BZ}}$ is a sum over wavevectors restricted to the first Brillouin Zone (BZ). 
The inverse relation
\begin{equation}\label{eq:inv op fourier transform}
    b^\dagger_{\mathbf{k}\boldsymbol{\delta}_{\ell}} = \dfrac{1}{\sqrt{N}} \sum_{\mathbf{R}} e^{-\mathrm{i}\mathbf{k}\cdot(\mathbf{R}+\boldsymbol{\delta}_{\ell})} c^\dagger_{\mathbf{R}+\boldsymbol{\delta}_{\ell}}
\end{equation}
is guaranteed by the orthogonality of the Fourier basis
\begin{equation}\label{eq:fourier orthogonality}
    \sum_{\mathbf{R}} e^{\mathrm{i}\mathbf{k}\cdot\mathbf{R}} = N\delta_{\mathbf{k}\mathbf{0}},
\end{equation}
for $\mathbf{k}$ restricted to the Brillouin zone. Using Eqs.~\eqref{eq:op fourier transform} and \eqref{eq:fourier orthogonality}, we rewrite the Hamiltonian (\ref{eq:tight-binding hamiltonian}) in momentum space as
\begin{equation}\label{eq:tight-binding hamiltonian momentum space}
    \mathcal{H}_{\text{TB}} \hspace{-0.1cm}=\hspace{-0.1cm} \sum_{\mathbf{k}}^{\text{BZ}} \hspace{-0.1cm} \sum_{\ell\ell'} \hspace{-0.1cm} \sum_{\boldsymbol{\delta}_{\ell}\boldsymbol{\delta}_{\ell'}} \hspace{-0.1cm} \sum_{\mathbf{R}} \hspace{-0.05cm} e^{\mathrm{i}\mathbf{k}\cdot(\mathbf{R}+\boldsymbol{\delta}_{\ell}-\boldsymbol{\delta}_{\ell'})} t(\mathbf{R} \hspace{-0.05cm} + \hspace{-0.05cm} \boldsymbol{\delta}_{\ell} \hspace{-0.05cm} - \hspace{-0.05cm} \boldsymbol{\delta}_{\ell'}) b^\dagger_{\mathbf{k}\boldsymbol{\delta}_{\ell}} b_{\mathbf{k}\boldsymbol{\delta}_{\ell'}},
\end{equation}
where $\mathbf{k}$ is the Bloch wavevector and $b^\dagger_{\mathbf{k}\boldsymbol{\delta}_{\ell}}$ is the transformed basis, indexed by the finite set of vectors $\boldsymbol{\delta}_{\ell}$.

We emphasize the importance of deriving a formulation for the Hamiltonian on a finite basis. This allows us to numerically construct and diagonalize Hamiltonian matrices. Here, we used the finite set of vectors $\boldsymbol{\delta}_{\ell}$ to label this basis, taking the periodicity of the infinite system into account. As shall be discussed in Sec.~\ref{sec.SI.transfer.integral}, the transfer integrals are position-dependent functions that decay exponentially with distance. Thus, the infinite sum over lattice vectors, $\sum_{\mathbf{R}}$, in Eq.~\eqref{eq:tight-binding hamiltonian momentum space} can be safely truncated.

\subsection{Transfer integral}\label{sec.SI.transfer.integral}

In the tight-binding picture, the transfer integrals $t(\mathbf{r})$ describe the energy parameters associated with the hoppings of electrons between different atomic sites. Here, one considers one orbital ($p_z$) per atomic site. Since the $p_z$ orbitals do not all align vertically due to the presence of two stacked graphene layers, geometrical spatial aspects of the $p_z$ orbital distribution must be considered. We closely follow the clever procedure used in Ref.~[\onlinecite{mirzakhani2020}] to handle this problem by considering that any interaction between pairs of $p_z$ orbitals can be decomposed into pure $\pi$- and $\sigma$-like bonds, as will be discussed as follows.

\begin{figure}[!t]
    \centering
    \includegraphics[width=1\columnwidth] {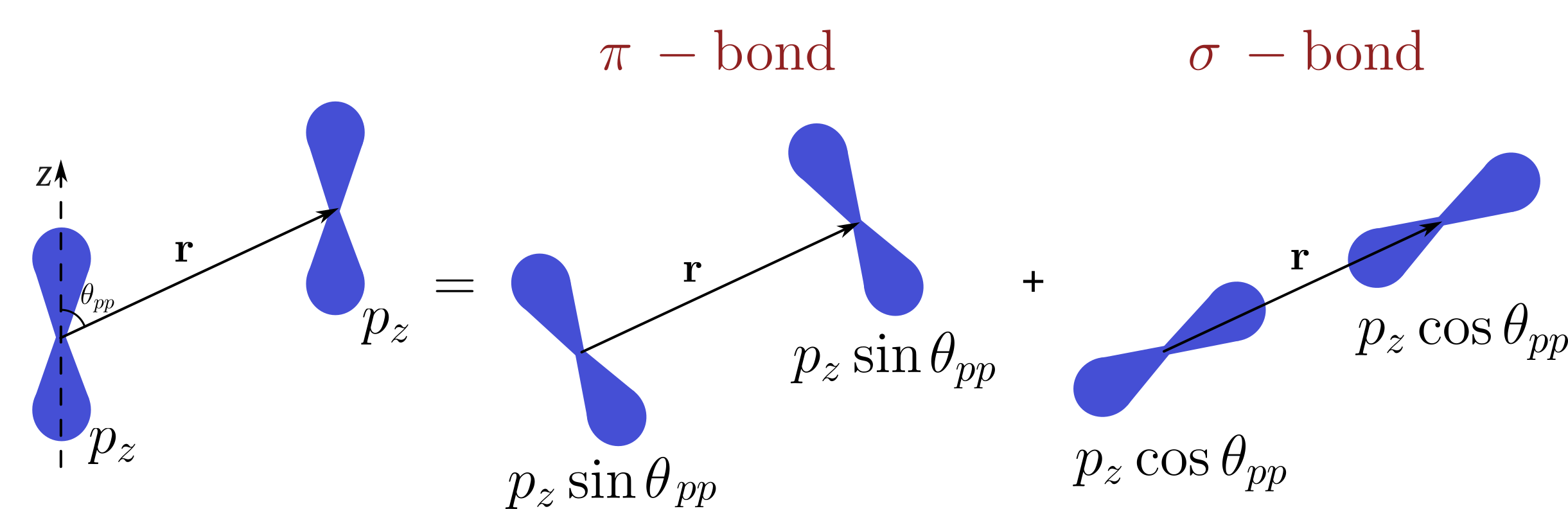}
    \caption{(Color online) Illustration of the interactions between a pair of $p_z$ orbitals, decomposed into $\pi$- and $\sigma$-like interactions. The lobes are projected onto a pair of axes, one parallel and one perpendicular to the vector $\mathbf{r}$ connecting the orbital centers.}
    \label{fig:pp}
\end{figure}

If $\mathbf{r}$ is the distance vector between two $p_z$ orbitals, then $\cos\theta_{pp}=(\mathbf{r}\cdot\mathbf{e}_z)/r$ is the cosine of the angle that $\mathbf{r}$ forms with the $z$ axis. Each $p_z$ orbital can be decomposed into a $p_z\cos\theta_{pp}$ component in the direction parallel to $\mathbf{r}$, and a $p_z\sin\theta_{pp}$ component in the orthogonal direction. The $p_z\cos\theta_{pp}$--$p_z\cos\theta_{pp}$ and the $p_z\sin\theta_{pp}$--$p_z\sin\theta_{pp}$ interactions resemble a $\sigma$-bond and a $\pi$-bond, respectively, as illustrated in Fig.~\ref{fig:pp}. Orthogonal interactions $p_z\cos\theta_{pp}$--$p_z\sin\theta_{pp}$ vanish due to the opposite signs of the $p_z$ orbital globes. Denoting the pure $\sigma$- and $\pi$-like transfer integrals by $V_{pp\sigma}(r)$ and $V_{pp\pi}(r)$, we can write the transfer integrals in terms of the Slater-Koster form as\cite{slater1954simplified, koshino2015electronic, moon2013optical}
\begin{equation}\label{eq.t.before}
    t(\mathbf{r}) = V_{pp\sigma}(r)\cos^2\theta_{pp} + V_{pp\pi}(r)\sin^2\theta_{pp}.
\end{equation}

Using the trigonometric identity $\sin^2\theta_{pp}=1-\cos^2\theta_{pp}$ and $\cos\theta_{pp}$ definition, Eq.~\eqref{eq.t.before} becomes \cite{mirzakhani2020}:
\begin{equation}\label{eq:transfer integral}
    t(\vb r)
    = 
    V_{pp\sigma}(r)\left(\dfrac{\vb{r}\cdot\vb e_z}{r}\right)^2 
    + V_{pp\pi}(r) \left[1-\left(\dfrac{\vb{r}\cdot\vb e_z}{r}\right)^2\right].
\end{equation}

Atomic orbitals have exponentially decaying tails far from their centers. For this reason, the pure $\sigma$ and $\pi$ transfer integrals are also assumed to decay exponentially. Thus, they are modeled as exponential functions with fitting parameters chosen to reflect the physical properties of the real system. The pure transfer integrals in Eq.~\eqref{eq:transfer integral} are fitted as
\begin{subequations}
\begin{align}
    V_{pp\pi}(r) &= V_{pp\pi}^0 \exp(-\dfrac{r - a_{cc}}{\delta_0}), \label{eq:pure pi hopping} \\
    V_{pp\sigma}(r) &= V_{pp\sigma}^0 \exp(-\dfrac{r - d_0}{\delta_0}), \label{eq:pure sigma hopping}
\end{align}
\end{subequations}
where $a_{cc}=a/\sqrt{3}\approx 1.42$ \AA\ is the carbon-carbon bond length and $d_0\approx 3.35$ \AA\ is the strain-free interlayer distance. The intralayer and interlayer nearest-neighbor hoppings are, respectively, given by
\begin{align}
    V_{pp\pi}(a_{cc}) &= V_{pp\pi}^0 \approx -2.7 \, \text{eV},\\
    V_{pp\sigma}(d_0) &= V_{pp\sigma}^0 \approx 0.48 \, \text{eV},
\end{align}
being set to agree with monolayer graphene and AB-stacked bilayer graphene band structures, respectively. The decay length $\delta_0=0.148 a$ is associated with the decay rates of the pure transfer integrals as
\begin{equation}
    \dfrac{1}{\delta_0}=-\dfrac{1}{V_{pp\sigma}(r)} \dfrac{\text{d} V_{pp\sigma}}{\text{d}r} (r)
    =-\dfrac{1}{V_{pp\pi}(r)} \dfrac{\text{d} V_{pp\pi}}{\text{d}r} (r).
\end{equation}
To truncate the sum $\sum_{\mathbf{R}}$ in Eq.~\eqref{eq:tight-binding hamiltonian momentum space} preserving a correct physical description of intra- and interlayer coupling in TBG, as mentioned in Sec.~\ref{sec:TBG tight-binding Hamiltonian}, we must identify the dominant intra- and interlayer hopping terms.
Following the approach of Ref.~[\onlinecite{mirzakhani2020}], we restrict intralayer hoppings to the nearest intralayer neighbors, and the interlayer ones, to atomic distances of $r \leq 4 a_{cc}$.

It is worth mentioning that applying any strain or pressure in a crystal modifies the vectors connecting lattice sites and changes the corresponding hopping parameters. Note that the transfer integral \eqref{eq:transfer integral} is a position-dependent term, implying hopping variations by lattice changes. Here, since one assumes rigid structures, TBG under pressure can be simply modeled by uniformly varying the interlayer distance.

\subsection{Electric-field effect inclusion into tight-binding Hamiltonian}

By applying an uniform electric field $\boldsymbol{\mathcal{E}}=\mathcal{E}\mathbf{e}_{z}$ perpendicular to TBG, a vertical electric potential $v(z)=v(0)-\mathcal{E}z$ is induced. We set the reference $v(d/2) = 0$, such that the potentials at the graphene layers are $v(0) = \mathcal{E}d/2$ and $v(d) = -\mathcal{E}d/2$. Thus, within the tight-binding approach, the bias voltage is included as an on-site potential given by the following diagonal term\cite{castro2010}
\begin{equation}\label{eq:hamiltonian electric field}
    \mathcal{H}_{V}
    = 
    \dfrac{V}{2} 
    \sum_{\mathbf{R}} \sum_{\ell \boldsymbol{\delta}_{\ell}} 
    s_{\ell}
    c^{\dagger}_{\mathbf{R} + \boldsymbol{\delta}_\ell} 
    c_{\mathbf{R} + \boldsymbol{\delta}_\ell}
    =
    \dfrac{V}{2}
    \sum_{\mathbf{k}}^{\text{BZ}} \sum_{\ell \boldsymbol{\delta}_{\ell}}
    s_{\ell}
    b^{\dagger}_{\mathbf{k} \boldsymbol{\delta}_\ell} 
    b_{\mathbf{k} \boldsymbol{\delta}_\ell},
\end{equation}
where $s_{\ell}=\delta_{\ell 1}-\delta_{\ell 2}$ and $V=e[v(0) - v(d)]=e\mathcal{E}d$ is the total electric potential energy difference between the adjacent graphene layers. The equivalence between the first and second right-hand-side terms in Eq.~\eqref{eq:hamiltonian electric field} is achieved using Eqs.~\eqref{eq:op fourier transform} and \eqref{eq:fourier orthogonality}.

\subsection{Total tight-binding TBG Hamiltonian and band structure calculations} \label{sec:H diag and band struc}

Combining Eqs.~\eqref{eq:tight-binding hamiltonian momentum space} and \eqref{eq:hamiltonian electric field}, we obtain the following total tight-binding TBG Hamiltonian for non-interacting $p_z$ electrons under an external perpendicular electric field
\begin{equation}\label{eq:H0}
    \mathcal{H}_0 = \mathcal{H}_{\text{TB}} + \mathcal{H}_{V} = \sum_{\mathbf{k}}^{\text{BZ}} \sum_{\ell\ell'} \sum_{\boldsymbol{\delta}_{\ell} \boldsymbol{\delta}_{\ell'}} h_{\boldsymbol{\delta}_{\ell} \boldsymbol{\delta}_{\ell'}} (\mathbf{k}) b^\dagger_{\mathbf{k}\boldsymbol{\delta}_{\ell}} b_{\mathbf{k}\boldsymbol{\delta}_{\ell'}},
\end{equation}
where we defined the matrix element
\begin{equation}\label{eq:matrix element}
    h_{\boldsymbol{\delta}_{\ell}\boldsymbol{\delta}_{\ell'}}(\mathbf{k}) = 
    \dfrac{V}{2} s_{\ell}\delta_{\ell\ell'}\delta_{\boldsymbol{\delta}_{\ell} \boldsymbol{\delta}_{\ell'}}
    +\sum_{\mathbf{R}} e^{\mathrm{i}\mathbf{k}\cdot(\mathbf{R}+\boldsymbol{\delta}_{\ell}-\boldsymbol{\delta}_{\ell'})} t(\mathbf{R}+\boldsymbol{\delta}_{\ell}-\boldsymbol{\delta}_{\ell'}).
\end{equation}

To diagonalize the Hamiltonian \eqref{eq:H0}, we introduce the Bloch operator
\begin{equation}\label{eq:bloch operator}
    a_{n\mathbf{k}}^\dagger = \sum_{\boldsymbol{\delta}_{\ell}} u_{n\boldsymbol{\delta}_{\ell}}(\mathbf{k}) b_{\mathbf{k}\boldsymbol{\delta}_{\ell}}^\dagger,
\end{equation}
where $n$ is the band label and the Bloch functions $u_{n\boldsymbol{\delta}_{\ell}}(\mathbf{k})$ form a orthonormal basis. 
The inverse relation
\begin{equation}\label{eq:bloch operator inverse}
    b^\dagger_{\mathbf{k}\boldsymbol{\delta}_{\ell}} = \sum_{n} u^*_{n\boldsymbol{\mathbf{\delta}}_{\ell}}(\mathbf{k}) a^\dagger_{n\mathbf{k}},
\end{equation}
is guaranteed by the orthonormality of the basis of functions $u_{n\boldsymbol{\delta}_{\ell}}(\mathbf{k})$ in both the indices $n$ and $\boldsymbol{\delta}_{\ell}$:
\begin{subequations}
\begin{align}
    \sum_{n} u_{n\boldsymbol{\delta}_{\ell}}^* (\mathbf{k}) u_{n\boldsymbol{\delta}_{\ell'}} (\mathbf{k}) &= \delta_{\boldsymbol{\delta}_{\ell} \boldsymbol{\delta}_{\ell'}}, \label{eq:u orthonormality delta}\\
    \sum_{\boldsymbol{\delta}_{\ell}} u_{n\boldsymbol{\delta}_{\ell}}^* (\mathbf{k}) u_{n'\boldsymbol{\delta}_{\ell}} (\mathbf{k}) &= \delta_{nn'}. \label{eq:u orthonormality n}
\end{align}
\end{subequations}
Substituting Eq.~\eqref{eq:bloch operator inverse} in Eq.~\eqref{eq:H0}, one gets
\begin{equation}\label{eq:H0 subs b to a}
    \mathcal{H}_0 = \sum_{\mathbf{k}}^{\text{BZ}} \sum_{\ell\ell'} \sum_{\boldsymbol{\delta}_{\ell}\boldsymbol{\delta}_{\ell'}} \sum_{nn'} h_{\boldsymbol{\delta}_{\ell}\boldsymbol{\delta}_{\ell'}}(\mathbf{k}) u^*_{n\boldsymbol{\delta}_{\ell}}(\mathbf{k}) u_{n'\boldsymbol{\delta}_{\ell'}}(\mathbf{k}) a^\dagger_{n\mathbf{k}} a_{n'\mathbf{k}}.
\end{equation}

Since $\mathcal{H}_0$ must be diagonal in the basis $a^\dagger_{n\mathbf{k}}$, Eq.~\eqref{eq:H0 subs b to a} can be split into the following pair of equations
\begin{subequations}
    \begin{align}
        &\sum_{\ell\ell'} \sum_{\boldsymbol{\delta}_{\ell}\boldsymbol{\delta}_{\ell'}} h_{\boldsymbol{\delta}_{\ell}\boldsymbol{\delta}_{\ell'}}(\mathbf{k}) u_{n\boldsymbol{\delta}_{\ell}}^*(\mathbf{k}) u_{n'\boldsymbol{\delta}_{\ell'}}(\mathbf{k}) = E_{n\mathbf{k}} \delta_{nn'}, \label{eq:eigenvalue problem}\\
        & \mathcal{H}_0 = \sum_{\mathbf{k}}^{\text{BZ}} \sum_{n} E_{n\mathbf{k}} a^\dagger_{n\mathbf{k}} a_{n\mathbf{k}}, \label{eq:H0 diagonal}
    \end{align}
\end{subequations}
where the eigenvalues $E_{n\mathbf{k}}$ describe the energy bands of the system. The Hamiltonian in the diagonal form \eqref{eq:H0 diagonal} will be important for further derivations in this work. To obtain the energy bands and Bloch functions, one should solve the eigenvalue problem \eqref{eq:eigenvalue problem}. For that, by using \eqref{eq:u orthonormality n}, we can rewrite Eq.~\eqref{eq:eigenvalue problem} in the standard form
\begin{equation}\label{eq:eigenvalue problem standard form}
    \sum_{\ell'\boldsymbol{\delta}_{\ell'}} h_{\boldsymbol{\delta}_{\ell}\boldsymbol{\delta}_{\ell'}}(\mathbf{k}) u_{n\boldsymbol{\delta}_{\ell'}}(\mathbf{k}) = E_{n\mathbf{k}} u_{n\boldsymbol{\delta}_{\ell}}(\mathbf{k}),
\end{equation}
for matrix elements indexed by $\boldsymbol{\delta}_{\ell}$ and $\boldsymbol{\delta}_{\ell'}$. To numerically solve Eq.~\eqref{eq:eigenvalue problem standard form}, explicit values for $\mathbf{k}$ must be sampled, as illustrated by green small dots in Fig.~\ref{fig:tbg_lattice}(b). For that, we sampled $\mathbf{k}$-points, discretizing the momentum space, in the hexagonal sampling $\mathbf{k}$ region highlighted in Fig.~\ref{fig:tbg_lattice}(b). Any valid reciprocal unit cell with the same area as the first Brillouin zone could be chosen, for instance, the rectangular sampling region illustrated with small gray symbols in Fig.~\ref{fig:tbg_lattice}(b), where we could assume a rectangular grid with $N_i$ sampling points in the $i=x',y'$ directions. Within such rectangular discretization, one has that the sampling points can be written explicitly as
\begin{align}\label{eq:k-point sampling rectangular}
    \mathbf{k}_{n_{x'}n_{y'}}^{(i)} &= \left(\dfrac{n_{x'}}{N_{x'}}\dfrac{\sqrt{3}}{2}\mathbf{G}_{2}\cdot\mathbf{e}_y + \dfrac{n_{y'}}{N_{y'}}\mathbf{G}_{2}\cdot\mathbf{e}_x\right) \mathbf{e}_x \nonumber \\
    & + \left(\dfrac{n_{y'}}{N_{y'}}\mathbf{G}_{2}\cdot\mathbf{e}_y - \dfrac{n_{x'}}{N_{x'}}\dfrac{\sqrt{3}}{2}\mathbf{G}_{2}\cdot\mathbf{e}_x\right) \mathbf{e}_y,
\end{align}
where $n_{i} = 0, 1, 2, \hdots, N_i - 1$. In this manner, the total number of sampling points $N_{\mathbf{k}}=N_{x'} N_{y'}$ will also act as the number of unit cells that constitute our material as well as directly related to the chosen convergence parameter since the larger the $N_{\mathbf{k}}$ parameter, more accuracy is expected. Instead of the equivalent discussed rectangular Brillouin zone, we assume the hexagonal Brillouin zone with a triangular grid to ensure the $C_6$ system symmetry, as also shown in Fig.~\ref{fig:tbg_lattice}.
The points of such a grid are described by the equation
\begin{align}\label{eq:k-point sampling hexagonal}
    \mathbf{k}_{n_1 n_2} &= \dfrac{(\mathbf{r}_{n_1 n_2}\cdot \mathbf{e}_x)(\mathbf{G}_2 \cdot \mathbf{e}_y) + (\mathbf{r}_{n_1 n_2} \cdot \mathbf{e}_y)(\mathbf{G}_2\cdot\mathbf{e}_x)}{(N_{h}-1)\sqrt{3}} \mathbf{e}_x\nonumber\\ & + \dfrac{(\mathbf{G}_2\cdot\mathbf{e}_y)(\mathbf{r}_{n_1n_2}\cdot\mathbf{e}_y) - (\mathbf{G}_2\cdot\mathbf{e}_x)(\mathbf{r}_{n_1n_2} \cdot \mathbf{e}_x)}{(N_{h}-1)\sqrt{3}}\mathbf{e}_y,
\end{align}
where $N_h$ is a parameter that regulates the number of points of the grid, and
\begin{eqnarray}
    \mathbf{r}_{n_1n_2} = n_1\mathbf{u}_1 + n_2\mathbf{u}_2,
\end{eqnarray}
where $\mathbf{u}_{\bfrac{1}{2}} = (\pm\sqrt{3}, 1)/2$ is a pair of auxiliary vectors, and $n_1,n_2$ is a pair of integers.
The points $\mathbf{r}_{n_1 n_2}$ that generate points $\mathbf{k}_{n_1 n_2}$ in the first Brillouin Zone are $\mathbf{r}^{(i)}_{n_1 n_2}$ ($i=1,2,3,4,5,6$), where $n_1+n_2<N_h$ and
\begin{subequations}
    \begin{equation}
        \mathbf{r}_{n_1n_2}^{(1)}  = \mathbf{r}_{n_1 n_2},\; 0\leq n_1,n_2 < N_h,
    \end{equation}
    \begin{equation}        
        \mathbf{r}_{n_1n_2}^{(2)}  = \mathbf{r}_{n_1+n_2,-n_2},\; 0\leq n_1 < N_h, 0<n_2<N_h,
    \end{equation}
    \begin{equation}        
        \mathbf{r}_{n_1n_2}^{(3)}  = \mathbf{r}_{n_2,-n_1-n_2},\; 0<n_1<N_h, 0\leq n_2<N_h,
    \end{equation}
    \begin{equation}        
        \mathbf{r}_{n_1n_2}^{(4)}  = \mathbf{r}_{-n_2,-n_1},\; 0\leq n_1<N_h, 0< n_2<N_h,
    \end{equation}
    \begin{equation}        
        \mathbf{r}_{n_1n_2}^{(5)}  = \mathbf{r}_{-n_2,n_1+n_2},\; 0\leq n_1<N_h, 0< n_2<N_h,
    \end{equation}
    \begin{equation}        
        \mathbf{r}_{n_1n_2}^{(6)}  = \mathbf{r}_{-n_1-n_2,n_2}, 0<n_1,n_2<N_h,
    \end{equation}
\end{subequations}
which sums up to a total of $N_{\mathbf{k}} = 3N_h(N_h-1) + 1$ grid points.
As will be shown later, it will be necessary to evaluate the single-particle wave-function $u_{n\boldsymbol{\delta}_\ell}(\mathbf{k})$ at points $\mathbf{k}$ in the six Brillouin Zones adjacent to the first Brillouin Zone.
However, we know that $u_{n\boldsymbol{\delta}_\ell}(\mathbf{k})$ is periodic in reciprocal space following the rule
\begin{equation}
    u_{n\boldsymbol{\delta}_{\ell}}(\mathbf{k}) = u_{n\boldsymbol{\delta}_{\ell}}(\mathbf{k}+m_1\mathbf{G}_1+m_2\mathbf{G}_2),
\end{equation}
for any pair of integers $m_1,m_2$.
Additionally, the relation
\begin{equation}\label{eq:equiv outside BZ}
    \mathbf{k}_{n_1n_2}+m_1\mathbf{G}_1+m_2\mathbf{G}_2 = \mathbf{k}_{n'_1n'_2}
\end{equation}
holds for
\begin{subequations}
    \begin{equation}
        n'_1 = n_1+(N_h-1)(m_1+m_2),
    \end{equation}
    \begin{equation}
        n'_2 = n_2+(N_h-1)(m_2-2m_1).
    \end{equation}
\end{subequations}
For our purposes, Eq.~\eqref{eq:equiv outside BZ} will be restricted to values of $m_1,m_2$ that shift the grid points $\mathbf{k}_{n_1n_2}$ in the first Brillouin Zone to the six Brillouin Zones adjacent to it, \textit{i.e.},
$(m_1,m_2)=(1,0),(0,1),(-1,0),(0,-1),(1,1),(-1,-1)$.

\section{Bandgap opening: Different twist angles}

Using the tight-binding model derived in the previous section, we investigate the interplay of the twist angle, electric bias, and vertical pressure in TBG. The latter is mimicked by varying the interlayer distance $d$ away from the equilibrium position $d_0=3.35$ \AA. For sufficiently small unit cells, the eigenvalue problem \eqref{eq:eigenvalue problem standard form} can be fully diagonalized numerically in momentum space for each sampled $\mathbf{k}$-point. In addition to stability reasons for dealing with rigid 2D structures, the mentioned numerical reason is the other why we will limit our discussion to large angles. For instance, results in Fig.~\textcolor{blue}{1} of the main text for $\theta(1,1) = 21.79^\circ$ and $\theta(1,6)=  46.8^\circ$ correspond to unit cells of the superlattice with $28$ and $76$ atoms, respectively.

\begin{figure}[t]
    \centering
    \includegraphics[width=1\columnwidth] {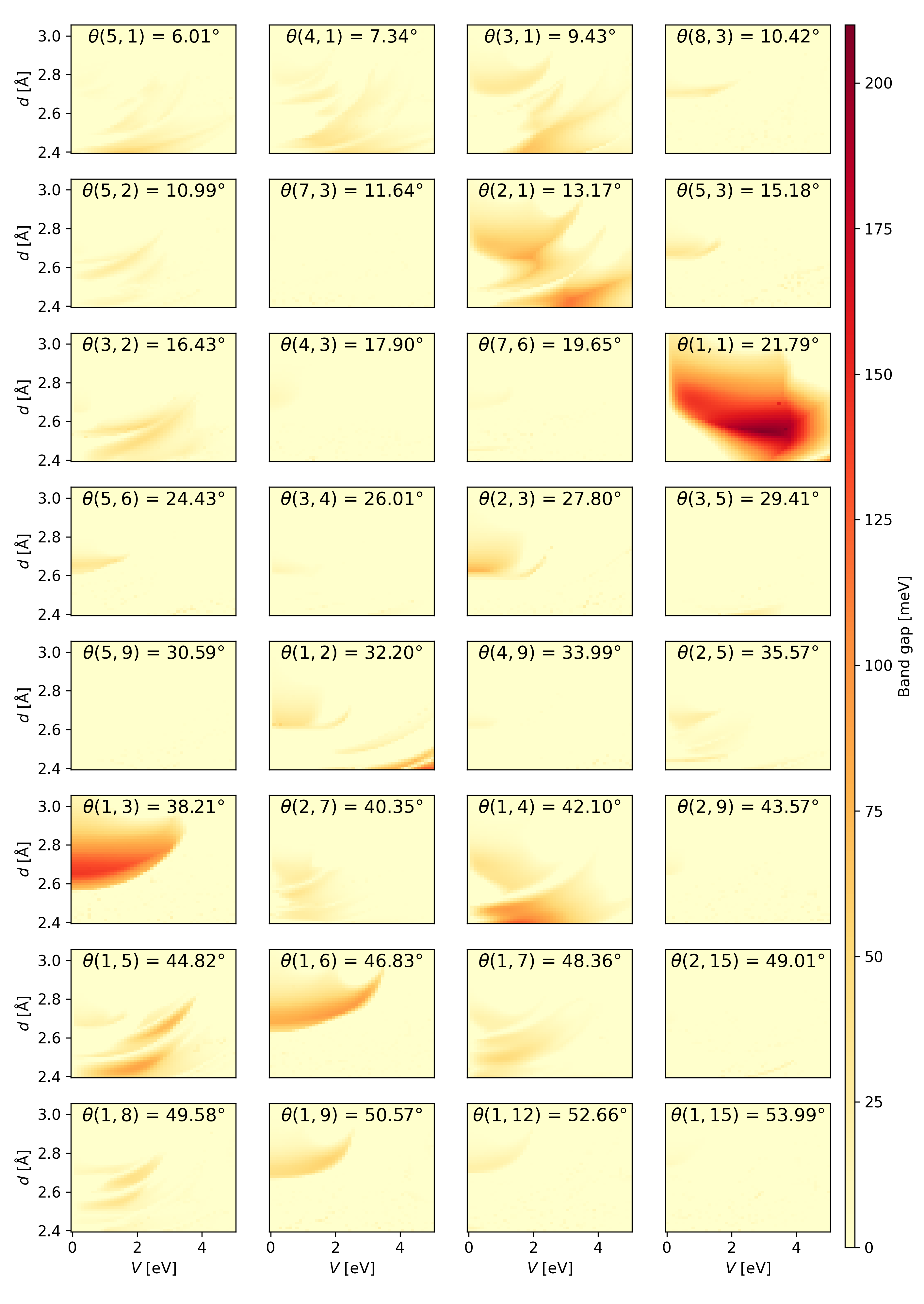}
    \caption{(Color online) Contour plots of the bandgap induced in TBG due to interlayer distance $d$ changes with respect to the equilibrium position $d_0$, mimicking the application of vertical pressure, and under a perpendicularly applied electric field with bias voltage $V$, for all twist angles that generate commensurate unit cells with less than 500 atoms.}
    \label{fig:gap search}
\end{figure}

In addition to Figs.~\textcolor{blue}{1(a)} and \textcolor{blue}{1(b)} and the corresponding discussion on the main text, here we present in Fig.~\ref{fig:gap search} the electronic bandgap of TBG for all the commensurable twist angles \eqref{eq:commensuration condition} that generate unit cells of less than 500 atoms, as a function of the bias potential and the interlayer distance. In general, one can notice: (i) highly twisting angle-dependent metallic-to-semiconductor transition induced solely by pressure or by an external electric field, or by the combination of them; and (ii) $\theta(1,3k)$ ($k\in\mathbb{Z}$) twist angles exhibit a particular gap tendency, increasing as a function of a decreasing in $d$, until some optimal value where the gap is maximum. This optimal point is brought closer to the equilibrium interlayer distance $(d=d_0)-$case as $V$ increases. This tunability to maximize the bandgap by applying an electric field is, experimentally speaking, more feasible to be performed \textit{in-situ}, demanding lower pressures that are more realistic to reach in experiments.

\section{Dielectric screening}\label{sec.S3}

In Sec.~\ref{sec.S2}, the electronic structure of single-particle excitations, considering non-interacting electrons, was described using the tight-binding formalism. To describe exciton formation, we must add many-body interactions to the model, \textit{i.e.}, electron-electron and electron-hole interactions. For this, in the current section, we shall derive analytically the static dielectric function of TBG, denoted here by $\epsilon$, which will describe how the system is polarized due to the application of an external electric field and how this polarization screens the electrostatic response of the system itself. In this work, the dielectric function is calculated in the context of the Random Phase Approximation (RPA) \cite{bohm1951collective, pines1952collective, bohm1953collective} for a periodic system \cite{adler1962,wiser1963}.

\subsection{Fourier transforms}

For a system of electrons perturbed by a dynamic external potential $\phi_{\text{ext}}$, the RPA states that the electrons will respond to $\phi_{\text{ext}}$ and a screening potential $\phi_{\text{scr}}$ induced by electrons themselves. Like any well-behaved 1D function, the time dependence of $\phi_{\text{ext}}$ and other relevant physical quantities can be Fourier transformed to the frequency domain using the convention
\begin{equation}
    \phi_{\text{ext}}(\mathbf{r}_{\parallel},z,t) = \int \dfrac{\text{d} \omega}{2\pi}\, e^{-\mathrm{i}\omega t} \phi_{\text{ext}}(\mathbf{r}_{\parallel},z,\omega),
\end{equation}
with inverse relation given by
\begin{equation}
    \phi_{\text{ext}}(\mathbf{r}_{\parallel},z,\omega) = \int \text{d} t\, e^{\mathrm{i}\omega t} \phi_{\text{ext}}(\mathbf{r}_{\parallel},z,t),
\end{equation}
where $\mathbf{r}_{\parallel}$ is an in-plane real space vector, $z$ is the out-of-plane coordinate, $t$ is time and $\omega$ is frequency. This gives the spectral composition of $\phi_{\text{ext}}$ in terms of pure harmonic components as a function of the oscillation frequency $\omega$. As will become evident in subsequent discussions, we will be solely concerned with the electronic response in the static regime, where $\omega$ goes to zero. Physically, this means that the external potential must be constant in time.

Similarly, the in-plane real space dependence of $\phi_{\text{ext}}$ and any other relevant one-body quantities are Fourier transformed to the momentum domain using the convention
\begin{equation}\label{eq:Fourier transform real space}
    \phi_{\text{ext}}(\mathbf{r}_{\parallel},z) = \dfrac{1}{\mathcal{S}} \sum_{\mathbf{k}} e^{\mathrm{i}\mathbf{k}\cdot\mathbf{r}_{\parallel}} \phi_{\text{ext}}(\mathbf{k},z),
\end{equation}
where $\mathcal{S}$ is the total surface area of the 2D material. In the case of TBG, we have $\mathcal{S}=N_{\mathbf{k}} \norm{\mathbf{L}_1\times\mathbf{L}_2}$.
The inverse relation is
\begin{equation}\label{eq:inverse Fourier transform real space}
    \phi_{\text{ext}}(\mathbf{k},z) = \int \text{d}\mathbf{r}_{\parallel}\, e^{-\mathrm{i}\mathbf{k}\cdot\mathbf{r}_{\parallel}} \phi_{\text{ext}}(\mathbf{r}_{\parallel},z).
\end{equation}

For ease of notation, we will use $\mathbf{r}=\mathbf{r}_\parallel$ and a superindex $\ell$ to indicate the $z$-dependence restricted to $z=0$ ($\ell=1$) and $z=d$ ($\ell=2$) from now on in Sec.~\ref{sec.S3}. Therefore, the dielectric function is defined as the function that satisfies
\begin{equation}\label{eq:dielectric function}
    \phi_{\text{ext}}^{\ell}(\mathbf{r}) = \int \text{d}\mathbf{r}' \epsilon^{\ell\ell'}(\mathbf{r},\mathbf{r}') \phi^{\ell'}(\mathbf{r}'),
\end{equation}
where $\phi=\phi_{\text{ext}}+\phi_{\text{scr}}$. The inverse dielectric function is defined according to the orthogonality relation
\begin{equation}\label{eq:dielectric orthogonality}
    \int \text{d}\mathbf{r}\, \epsilon^{\ell\ell'}(\mathbf{r}_1,\mathbf{r}) {\epsilon^{-1}}^{\ell\ell'}(\mathbf{r},\mathbf{r}_2) = \delta_{\ell\ell'}\delta(\mathbf{r}_1-\mathbf{r}_2),
\end{equation}
which allows us to rewrite Eq.~\eqref{eq:dielectric function} as 
\begin{equation}\label{eq:inverse dielectric function}
    \phi^{\ell}(\mathbf{r}) = \int \text{d}\mathbf{r}' {\epsilon^{-1}}^{\ell\ell'}(\mathbf{r},\mathbf{r}') \phi^{\ell}_{\text{ext}}(\mathbf{r}').
\end{equation}

The dielectric function (and analogously, its inverse) is Fourier transformed using the convention
\begin{equation}\label{eq:Fourier transform dielectric}
    \epsilon^{\ell\ell'}(\mathbf{r},\mathbf{r}') = \dfrac{1}{\mathcal{S}^2} \sum_{\mathbf{k}\mathbf{k}'} e^{\mathrm{i}\mathbf{k}\cdot\mathbf{r}} \epsilon^{\ell\ell'}(\mathbf{k},\mathbf{k}') e^{-\mathrm{i}\mathbf{k}'\cdot\mathbf{r}'},
\end{equation}
whose inverse is
\begin{equation}\label{eq:inverse Fourier transform dielectric}
    \epsilon^{\ell\ell'}(\mathbf{k},\mathbf{k}') = \int \text{d}\mathbf{r} \int \text{d}\mathbf{r}' e^{-\mathrm{i}\mathbf{k}\cdot\mathbf{r}} \epsilon^{\ell\ell'}(\mathbf{r},\mathbf{r}') e^{\mathrm{i}\mathbf{k}'\cdot\mathbf{r}'}.
\end{equation}
The orthogonality relation becomes
\begin{equation}\label{eq:dielectric orthogonality momentum}
    \sum_{\mathbf{k}} \epsilon^{\ell\ell'}(\mathbf{k}_1,\mathbf{k}) {\epsilon^{-1}}^{\ell\ell'}(\mathbf{k},\mathbf{k}_2) = \delta_{\ell\ell'}\delta_{\mathbf{k}_1\mathbf{k}_2}.
\end{equation}
Using the Fourier transform conventions \eqref{eq:Fourier transform real space}-\eqref{eq:Fourier transform dielectric}, we can take Eqs.~\eqref{eq:dielectric function} and \eqref{eq:inverse dielectric function} to momentum space, such as
\begin{subequations}
    \begin{align}
        \phi^{\ell}_{\text{ext}}(\mathbf{k}) &= \sum_{\mathbf{k}'} \epsilon(\mathbf{k},\mathbf{k}') \phi^{\ell}(\mathbf{k}'), \label{eq:dielectric function momentum}\\       
        \phi^{\ell}(\mathbf{k}) &= \sum_{\mathbf{k}'} \epsilon^{-1}(\mathbf{k},\mathbf{k}') \phi^{\ell}_{\text{ext}}(\mathbf{k}'). \label{eq:inverse dielectric function momentum}
    \end{align}
\end{subequations}
Equations \eqref{eq:dielectric function momentum} and \eqref{eq:inverse dielectric function momentum} connect the external and total potentials in momentum space through the direct and inverse dielectric functions. They will be the starting point for the discussions of the next subsection [Sec.~\ref{sec:dielectric function formalism}], where one establishes the formalism necessary to perform dielectric function calculations in momentum space.

\subsection{Formalism} \label{sec:dielectric function formalism}

In this section, we present a procedure derived to express the dielectric function solely in terms of eigenenergies $E_{n\mathbf{k}}$ and the overlap of the wavefunctions $u_{n\boldsymbol{\delta}}(\mathbf{k})$ of the unperturbed system, described by the Hamiltonian $\mathcal{H}_0$ present in Sec.~\ref{sec.S2} and in the main text. Such theoretical derivation is based on the works of Refs.~[\onlinecite{adler1962, wiser1963}] for layered 2D materials. We will consider that the potential at $\mathbf{k}$ is only affected by $\mathbf{k}'=\mathbf{k}$ and its periodical repetitions, which can be included in our model by restricting the dielectric function to
\begin{equation}\label{eq:screening contributions momentum space}
    \epsilon^{\ell\ell'}(\mathbf{k},\mathbf{k}') = \sum_{\mathbf{G}_1} \epsilon^{\ell\ell'}(\mathbf{k},\mathbf{k}') \delta_{\mathbf{k}+\mathbf{G}_1,\mathbf{k}'},
\end{equation}
the remaining contributions are assumed to average out to zero, which is well justified in the RPA context. In fact, we could go a step further and restrict the screening contributions to the dominant term $\mathbf{G}_1=0$ ($\mathbf{k}'=\mathbf{k}$) only, meaning that the dielectric function would be approximated to a local function in momentum space. However, the formulation \eqref{eq:screening contributions momentum space} allows us to be a bit broader and inspect how the terms $\mathbf{G}_1\neq 0$ affect the screening if necessary. Expanding $\mathbf{k}$ and $\mathbf{k}'$ as $\mathbf{k}=\mathbf{q}+\mathbf{G}$ and $\mathbf{k}'=\mathbf{q}'+\mathbf{G}'$, respectively, with $\mathbf{q}$ and $\mathbf{q}'$ wavevectors restricted to the Brillouin zone, and $\mathbf{G}$ and $\mathbf{G}'$ being the reciprocal lattice vectors, Eq.~\eqref{eq:screening contributions momentum space} can be rewritten as
\begin{align}
    \epsilon^{\ell\ell'}\hspace{-0.05cm}(\mathbf{q}\hspace{-0.075cm}+\hspace{-0.075cm}\mathbf{G},\mathbf{q}'\hspace{-0.075cm}+\hspace{-0.075cm}\mathbf{G}') &\hspace{-0.1cm}=\hspace{-0.125cm} \sum_{\mathbf{G}_1}\hspace{-0.025cm} \epsilon^{\ell\ell'}\hspace{-0.025cm}(\mathbf{q}\hspace{-0.075cm}+\hspace{-0.075cm}\mathbf{G},\mathbf{q}'\hspace{-0.075cm}+\hspace{-0.075cm}\mathbf{G}') \delta_{\mathbf{q}+\mathbf{G}+\mathbf{G}_1,\mathbf{q}'+\mathbf{G}'}\nonumber\\
    &\hspace{-0.1cm}= \hspace{-0.1cm}\sum_{\mathbf{G}'_1}\hspace{-0.025cm} \epsilon^{\ell\ell'} \hspace{-0.025cm}(\mathbf{q}\hspace{-0.075cm}+\hspace{-0.075cm}\mathbf{G},\mathbf{q}'\hspace{-0.075cm}+\hspace{-0.075cm}\mathbf{G}') \delta_{\mathbf{q}+\mathbf{G}'_1,\mathbf{q}'}\nonumber \\
    &\hspace{-0.1cm}=\hspace{-0.1cm} \epsilon^{\ell\ell'}_{\mathbf{G}\mathbf{G}'}(\mathbf{q}) \delta_{\mathbf{q},\mathbf{q}'},
\end{align}
where we defined $\epsilon^{\ell\ell'}_{\mathbf{G}\mathbf{G}'}(\mathbf{q}) = \epsilon^{\ell\ell'}(\mathbf{q}+\mathbf{G},\mathbf{q}+\mathbf{G}')$. Using this result, we rewrite Eq.~\eqref{eq:dielectric orthogonality momentum} as
\begin{equation}
    \sum_{\ell\mathbf{G}} \epsilon^{\ell_1\ell}_{\mathbf{G}_1\mathbf{G}}(\mathbf{q}) {\epsilon^{-1}}^{\ell\ell_2}_{\mathbf{G}\mathbf{G}_2}(\mathbf{q}) = \delta_{\ell_1\ell_2}\delta_{\mathbf{G}_1\mathbf{G}_2}, \label{eq:dielectric orthogonality final}
\end{equation}
and, subsequently, Eqs.~\eqref{eq:dielectric function momentum} and \eqref{eq:inverse dielectric function momentum} become
\begin{subequations}
    \begin{align}
        \phi_{\text{ext}}^{\ell}(\mathbf{q}+\mathbf{G}) &= \sum_{\ell'\mathbf{G}'} \epsilon^{\ell\ell'}_{\mathbf{G}\mathbf{G}'}(\mathbf{q})\phi^{\ell'}(\mathbf{q}+\mathbf{G}'), \label{eq:dielectric function final}\\
        \phi^{\ell}(\mathbf{q}+\mathbf{G}) &= \sum_{\ell'\mathbf{G}'} {\epsilon^{-1}}^{\ell\ell'}_{\mathbf{G}\mathbf{G}'}(\mathbf{q})\phi_{\text{ext}}^{\ell'}(\mathbf{q}+\mathbf{G}'). \label{eq:inverse dielectric function final}
    \end{align}
\end{subequations}

When excitons are formed in TBG, the electron-hole interaction is screened by its surroundings. This effect must be taken into account to describe the excitonic spectrum accurately. As mentioned previously, in this section, we present a formalism adapted from Refs.~[\onlinecite{adler1962, wiser1963}] to take the screening effect into account through a description based on the linear electronic response of the $p_z$ TBG electrons in light of the RPA. The electronic distribution fluctuations around the ground state are associated with a certain induced potential $\phi_{\text{ind}}$. To study how excitons are formed in this system, instead of obtaining the exact form of $\phi_{\text{ext}}$, we will, however, approach this problem more cleverly by associating a dielectric function that will establish a direct relation between $\phi_{\text{ext}}$ and the total potential $\phi = \phi_{\text{ext}} + \phi_{\text{ind}}$. 

Considering each carbon site of the TBG as a point charge, such that variations in charge density due to fluctuations near the Fermi level can be written as
\begin{equation}\label{eq:charge density}
    \sigma_1(\mathbf{r})\delta(z) + \sigma_2(\mathbf{r})\delta(z-d),
\end{equation}
where $\sigma_\ell(\mathbf{r})$ is the surface charge density fluctuation of layer $\ell$ and $\delta(z)$ is the delta function. The induced potential is obtained by the solution of the Poisson equation for the charge density, given by Eq.~\eqref{eq:charge density}, such that
\begin{equation}\label{eq.poisson.charge.density}
    \nabla^2 \phi_{\text{ind}}(\mathbf{r},z) = -\dfrac{1}{\epsilon_0} \left[ \sigma_1(\mathbf{r})\delta(z) + \sigma_2(\mathbf{r})\delta(z-d) \right].
\end{equation}
By Fourier transform Eq.~\eqref{eq.poisson.charge.density}, one obtains the following solution in momentum space
\begin{equation}\label{eq:induced potential}
    \phi_{\text{ind}}^{\ell}(\mathbf{k}) = \sum_{\ell'} X^{\ell\ell'}(k) \sigma_{\ell'}(\mathbf{k}),
\end{equation}
where $X^{\ell\ell'}$ is a term associated to the Coulomb potential without screening
\begin{equation}\label{eq:coulomb potential momentum space}
    X^{\ell\ell'}(k) = \dfrac{\delta_{\ell\ell'} + (1-\delta_{\ell\ell'}) e^{-kd}}{2\epsilon_0 k},
\end{equation}
denominated bare Coulomb potential. This potential acts on the system through the one-body Hamiltonian
\begin{align}\label{eq:Hind}
    \mathcal{H}_{\text{ind}} &= -e \sum_{\ell\boldsymbol{\delta}_{\ell}} \sum_{\mathbf{R}} \phi^\ell(\mathbf{R}+\boldsymbol{\delta}_{\ell}) c^\dagger_{\mathbf{R}+\boldsymbol{\delta}_{\ell}} c_{\mathbf{R}+\boldsymbol{\delta}_{\ell}} \nonumber \\ 
    &= -\dfrac{e}{\mathcal{S}} \sum_{\mathbf{q}}^{\text{BZ}} \sum_{\ell} \sum_{\mathbf{G}} \phi^\ell(\mathbf{q}+\mathbf{G}) \nonumber\\
    & \times \sum_{\mathbf{k}}^{\text{BZ}} \sum_{nn'} M^{nn'}_{\ell}(\mathbf{k},\mathbf{q},\mathbf{G}) a^\dagger_{n\mathbf{k}} a_{n'\mathbf{k}+\mathbf{q}},
\end{align}
where we defined the overlap term
\begin{equation}\label{eq:M term}
    M^{nn'}_{\ell}(\mathbf{k},\mathbf{q},\mathbf{G}) = \sum_{\boldsymbol{\delta}_{\ell}} u^*_{n\boldsymbol{\delta}_{\ell}}(\mathbf{k}) e^{\mathrm{i}\mathbf{G}\cdot \boldsymbol{\delta}_{\ell}} u_{n'\boldsymbol{\delta}_{\ell}}(\mathbf{k}+\mathbf{q}).
\end{equation}

From the Liouville equation
\begin{equation}\label{eq:Liouville equation}
    -\mathrm{i}\hbar\partial_t \rho = [\mathcal{H},\rho],
\end{equation}
that describes the evolution of a quantum system of particles in terms of the distribution function $\rho$ in phase space associated with a Hamiltonian $\mathcal{H}$. For the unperturbed system, we have $\mathcal{H}=\mathcal{H}_0$ and $\rho=\rho_0$, where $\rho_0$ is the well-known distribution of independent electrons' system, which obeys the Pauli exclusion principle. Thus, we can write
\begin{equation}\label{eq:unperturbed distribution}
    \rho_0|n\mathbf{k}\rangle = f(E_{n\mathbf{k}}) |n\mathbf{k}\rangle,
\end{equation}
where $f(E)$ is the Fermi-Dirac distribution
\begin{equation}\label{eq:Fermi-Dirac distribution}
    f(E) = \dfrac{1}{\exp\{(E - E_{F})/k_B T\} + 1},
\end{equation}
with $k_B$ being the Boltzmann constant, $T$ the temperature, and $E_F$ the Fermi energy. 

The unperturbed distribution $\rho_0$ commutes with $\mathcal{H}_0$ since they are simultaneously diagonalized by the basis of Bloch states $|n\mathbf{k}\rangle$, as shown in Eqs.~\eqref{eq:H0 diagonal} and \eqref{eq:unperturbed distribution}. For this reason, the unperturbed distribution remains static, and Eq.~\eqref{eq:Liouville equation} can be simplified to
\begin{equation}\label{eq:unperturbed liouville}
    -\mathrm{i}\hbar\partial_t \rho_0 = [\mathcal{H}_0,\rho_0] = 0.
\end{equation}

Now we set $\mathcal{H}=\mathcal{H}_0 + \mathcal{H}_{\text{ind}}$ and consider fluctuations $\rho_{\text{ind}}$ around the distribution $\rho_0$ of the unperturbed system $\mathcal{H}_0$, whose eigenstates are $\ket{n\mathbf{k}} = a^\dagger_{n\mathbf{k}} \ket{0}$. Thus, $\rho=\rho_0 + \rho_{\text{ind}}$. On the other hand, $\rho_{\text{ind}}$ is closely related with fluctuations in the charge density, such as
\begin{equation}\label{eq:sigma-rho link real space}
    \sigma_\ell(\mathbf{r}) = -e \sum_{\boldsymbol{\delta}_{\ell}\mathbf{R}} \delta(\mathbf{r}-\mathbf{R}-\boldsymbol{\delta}_{\ell}) \langle \mathbf{R}+\boldsymbol{\delta}_{\ell}|\rho_{\text{ind}}|\mathbf{R}+\boldsymbol{\delta}_{\ell}\rangle,
\end{equation}
which is Fourier transformed to
\begin{align}\label{eq:sigma-rho link rec space}
    \sigma_{\ell}(\mathbf{q}+\mathbf{G}) &= -e \sum_{\mathbf{k}}^{\text{BZ}} \sum_{nn'} \left(M^{nn'}_{\ell}(\mathbf{k},\mathbf{q},\mathbf{G})\right)^*\nonumber\\
    &\times \langle n\mathbf{k}|\rho_{\text{ind}}|n' \mathbf{k}+\mathbf{q} \rangle.
\end{align}

To explicit the matrix elements of $\rho_{\text{ind}}$ in the basis $\ket{n\mathbf{k}}$, we recognize that, as a first-order approximation, the differential equation \eqref{eq:Liouville equation} can be linearized by setting $\rho_{\text{ind}}$ proportional to the perturbation $\mathcal{H}_{\text{ind}}$. Under this consideration, the commutator $[\mathcal{H}_{\text{ind}},\rho_{\text{ind}}]$ will vanish, as well as $[\mathcal{H}_0,\rho_0]$ and the time derivative $\partial_t \rho_0$ from Eq.~\eqref{eq:unperturbed liouville}. Thus, one obtains
\begin{equation}
    -\mathrm{i}\hbar\partial_t \rho_{\text{ind}} = [\mathcal{H}_0,\rho_{\text{ind}}] + [\mathcal{H}_{\text{ind}},\rho_0].
\end{equation}
The final step is to recognize that, in the static approximation, $\rho_{\text{ind}}$ will vary slowly. Therefore, setting $\partial_t \rho_{\text{ind}} \approx 0$, it gives
\begin{equation}\label{eq:linearized distribution}
    \langle n\mathbf{k}| \rho_{\text{ind}}|n'\mathbf{k}+\mathbf{q}\rangle \hspace{-0.05cm}=\hspace{-0.05cm} \dfrac{f(E_{n\mathbf{k}}) \hspace{-0.075cm}-\hspace{-0.075cm} f(E_{n'\mathbf{k}+\mathbf{q}})}{E_{n\mathbf{k}} \hspace{-0.075cm}-\hspace{-0.075cm} E_{n'\mathbf{k}+\mathbf{q}}} \langle n\mathbf{k}|\mathcal{H}_{\text{ind}}|n'\mathbf{k}+\mathbf{q}\rangle.
\end{equation}

Replacing Eq.~\eqref{eq:linearized distribution} into Eq.~\eqref{eq:sigma-rho link rec space}, then writing down explicitly the matrix elements $\bra{n\mathbf{k}}\mathcal{H}_{\text{ind}}\ket{n'\mathbf{k}+\mathbf{q}}$ and using Eq.~\eqref{eq:Hind}, one gets
\begin{align}\label{eq:sigma-rho link rec space final}
    \sigma_{\ell}(\mathbf{q}+\mathbf{G}) &= \dfrac{e^2}{\mathcal{S}} \sum_{\mathbf{k}}^{\text{BZ}} \sum_{nn'} \left(M^{nn'}_{\ell}(\mathbf{k},\mathbf{q},\mathbf{G})\right)^*\nonumber\\
    & \times\dfrac{f(E_{n\mathbf{k}}) - f(E_{n'\mathbf{k}+\mathbf{q}})}{E_{n\mathbf{k}} - E_{n'\mathbf{k}+\mathbf{q}}} \nonumber \\
    & \times \sum_{\ell'} \sum_{\mathbf{G}'} \phi^{\ell'}(\mathbf{q}+\mathbf{G}') M^{nn'}_{\ell'}(\mathbf{k},\mathbf{q},\mathbf{G}').
\end{align}

Merging Eqs.~\eqref{eq:induced potential} and \eqref{eq:sigma-rho link rec space final}, and rearranging some terms, it results in
\begin{align}\label{eq:glue}
    \phi^{\ell}_{\text{ind}}(\mathbf{q}+\mathbf{G}) &= \dfrac{e^2}{\mathcal{S}} \sum_{\ell'\mathbf{G}'} \phi^{\ell'}(\mathbf{q}+\mathbf{G}') \sum_{\ell''} X^{\ell\ell''}(|\mathbf{q}+\mathbf{G}|) \nonumber \\ 
    & \times \sum_{nn'} \sum_{\mathbf{k}}^{\text{BZ}} \left(M^{nn'}_{\ell''}(\mathbf{k},\mathbf{q},\mathbf{G})\right)^*\nonumber\\  
    & \times \dfrac{f(E_{n\mathbf{k}}) - f(E_{n'\mathbf{k}+\mathbf{q}})}{E_{n\mathbf{k}} - E_{n'\mathbf{k}+\mathbf{q}}}
     M^{nn'}_{\ell'}(\mathbf{k},\mathbf{q},\mathbf{G}'),
\end{align}
which establishes a relation between $\phi_{\text{ind}}$ and $\phi$ solely in terms of the energies $E_{n\mathbf{k}}$ and the overlap of wavefunctions $u_{n\mathbf{k}}$, both of them associated with the unperturbed system $\mathcal{H}_0$. Putting this aside for a moment, we recall that $\phi = \phi_{\text{ext}} + \phi_{\text{ind}}$ and use the dielectric function definition \eqref{eq:dielectric function final} to derive another relation between $\phi_{\text{ind}}$ and $\phi$:
\begin{equation}\label{eq:induced potential from dielectric function definition}
    \phi^\ell_{\text{ind}}(\mathbf{q}+\mathbf{G})=\sum_{\ell'\mathbf{G}'} \phi^{\ell'}(\mathbf{q}+\mathbf{G}') \left[\delta_{\ell\ell'}\delta_{\mathbf{G}\mathbf{G'}} - \epsilon^{\ell\ell'}_{\mathbf{G}\mathbf{G}'}(\mathbf{q}) \right].
\end{equation}

Comparing Eqs.~\eqref{eq:glue} and \eqref{eq:induced potential from dielectric function definition}, it yields the following expression for the dielectric function
\begin{align}\label{eq:dielectric function calc}
    \epsilon^{\ell\ell'}_{\mathbf{G}\mathbf{G}'}(\mathbf{q}) &= \delta_{\ell\ell'}\delta_{\mathbf{G}\mathbf{G}'} - \dfrac{e^2}{\mathcal{S}} \sum_{\ell''} X^{\ell\ell''}(|\mathbf{q}+\mathbf{G}|) \nonumber \\
    & \times \sum_{n n'} \sum_{\mathbf{k}}^{\text{BZ}} \left(M^{n n'}_{\ell''}(\mathbf{k},\mathbf{q},\mathbf{G})\right)^*\nonumber\\
    & \times \dfrac{f(E_{n\mathbf{k}}) - f(E_{n' \mathbf{k}+\mathbf{q}}) }{E_{n\mathbf{k}} - E_{n' \mathbf{k}+\mathbf{q}}}M^{n n'}_{\ell'}(\mathbf{k},\mathbf{q},\mathbf{G}'),
\end{align}
which can be seen as a set of matrix elements indexed by $\{\ell,\mathbf{G}\}$ and $\{\ell',\mathbf{G}'\}$. In the limit of zero temperature, we can rewrite Eq.~\eqref{eq:dielectric function calc} as
\begin{align}\label{eq:dielectric function calc final}
    &\epsilon^{\ell\ell'}_{\mathbf{G}\mathbf{G}'}(\mathbf{q}) = \delta_{\ell\ell'}\delta_{\mathbf{G}\mathbf{G}'} + \dfrac{e^2}{\mathcal{S}} \sum_{\ell''} X^{\ell\ell''}(|\mathbf{q}+\mathbf{G}|) \nonumber \\ 
    & \times \sum_{n_c n_v} \sum_{\mathbf{k}}^{\text{BZ}} \left[\dfrac{\left(M^{n_c n_v}_{\ell''}(\mathbf{k},\mathbf{q},\mathbf{G})\right)^* M^{n_c n_v}_{\ell'}(\mathbf{k},\mathbf{q},\mathbf{G}')}{E_{n_c \mathbf{k}} - E_{n_v \mathbf{k}+\mathbf{q}}} \right. \nonumber \\ 
    & \left. + \dfrac{\left(M^{n_v n_c}_{\ell''}(\mathbf{k},\mathbf{q},\mathbf{G})\right)^* M^{n_v n_c}_{\ell'}(\mathbf{k},\mathbf{q},\mathbf{G}')}{E_{n_c \mathbf{k}+\mathbf{q}} - E_{n_v \mathbf{k}}} \right],
\end{align}
where $n_c$ ($n_v$) is a band index that only sums over conduction (valence) bands.

At last, we emphasize that for $\mathbf{q}$ of the small norm, \textit{i.e.}, within the long-wavelength regime, the dominant term $\epsilon^{\ell\ell'}_{\mathbf{0}\mathbf{0}}$ can be simplified in light of the Rytova-Keldysh potential \cite{rytova1967, keldysh1979} for 2D systems. In this case, it takes the form
\begin{equation}\label{eq:eps linear}
    \epsilon^{\ell\ell'}_{\mathbf{G}\mathbf{G}'}(\mathbf{q})\approx (\delta_{\ell\ell'} + r_0^{\ell\ell'} q)\delta_{\mathbf{G}\mathbf{0}}\delta_{\mathbf{G}'\mathbf{0}},
\end{equation}
where $r_0^{\ell\ell'}$ are the screening lengths.

\subsection{Dielectric function color plot}

\begin{figure}[t]
    \centering
    \includegraphics[width=\linewidth]{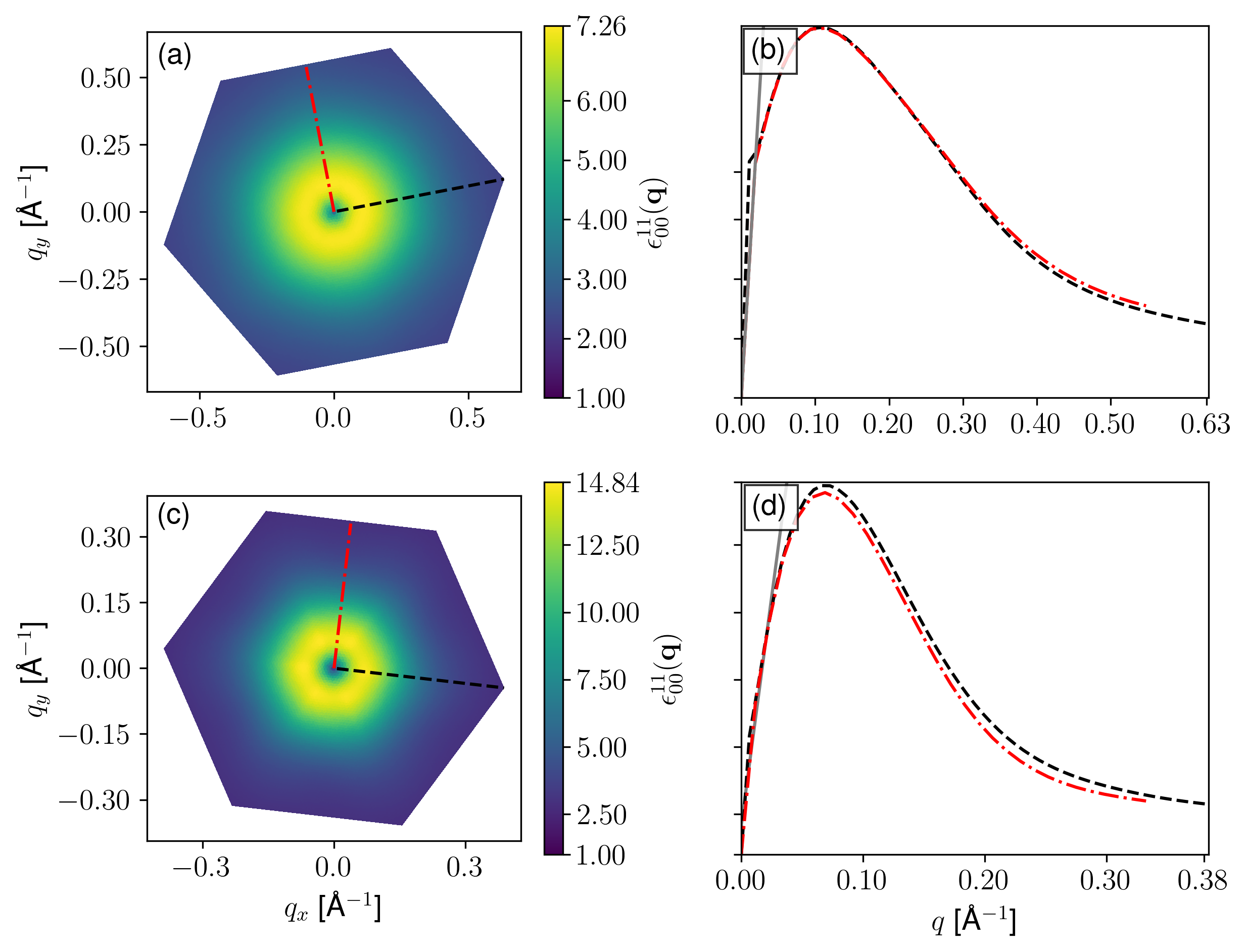}
    \caption{
        (Color online)
        Dominant term ($\ell=\ell'=0$, $\mathbf{G}=\mathbf{G}'=\mathbf{0}$) of the dielectric function in momentum space [see Eq.~\eqref{eq:dielectric function}] as a function of the transferred momentum $\mathbf{q}$, calculated in the hexagonal unit cell as shown by the green symbols inside the first Brillouin zone in Fig.~\ref{fig:tbg_lattice}(b), for a TBG under bias voltage $V=3$ eV, and twist angles and interlayer distances of (a) $\theta(1,1)= 21.79^\circ$ and $d=2.6$ \AA, and (c) $\theta(1,6) = 46.83^\circ$ and $d=2.8$ \AA. A pronounced ring-like peak is observed in the dielectric function around $(q_x,q_y)=(0,0)$. Cross-sections (red and black curves) of the dielectric function's contour plots (a) and (c) are depicted in (b) and (d), respectively, showing linear fit (gray solid) curves for small $|\mathbf{q}|$.}
    \label{fig:dielectric function}
\end{figure}

Figure~\ref{fig:dielectric function} presents color plots of the dielectric function calculated using Eq.~\eqref{eq:dielectric function calc final} restricted to the reciprocal lattice vectors $\mathbf{G}=\mathbf{G}'=\mathbf{0}$ and intralayer contribution $\ell=\ell'=0$, for biased TBG system with gate potential $V=3$ eV, taking the twist angle and interlayer distance as (a, b) $\theta(1,1)= 21.79^\circ$ and $d=2.6$ \AA, and (c, d) $\theta(1,6) = 46.83^\circ$ and $d=2.8$ \AA. From the contour plots \ref{fig:dielectric function}(a) and \ref{fig:dielectric function}(c), one can observe an almost circularly symmetric ring-like peak for small transferred momenta, exhibiting a linear increasing slope with $||\mathbf{q}||$, as emphasized by the gray solid curve fitting plot of the dielectric function's cross-sections in panels \ref{fig:dielectric function}(b) and \ref{fig:dielectric function}(d). Far from the $[\mathbf{q}=(0,0)]$--region, the dielectric matrix $\epsilon^{00}_{\mathbf{0}\mathbf{0}}(\mathbf{q})$ manifests qualitatively similar without major fluctuations in its magnitude. Moreover, one verifies that the ring-like peak value oscillates with $\angle\mathbf{q}$. Given that $\mathbf{q}=(0,q_{y'}^{\text{max}})$ and $\mathbf{q}=(q_{x'}^{\text{max}},0)$ cases maximize the dielectric function at fixed momentum $q_{x'}=0$ and $q_{y'}=0$ directions, respectively, where the ${(')}$--index indicates the rotated momenta directions, we can estimate about the isotropic or anisotropic aspect of the dielectric function. For instance, for $\theta(1,6) = 46.83^\circ$ shown in panels \ref{fig:dielectric function}(c) and \ref{fig:dielectric function}(d), one has the ratios $\epsilon^{00}_{\mathbf{0}\mathbf{0}}(0,q_{y'}^{max}) / \epsilon^{00}_{\mathbf{0}\mathbf{0}}(q_{x'}^{\text{max}},0) \approx 1.08$ and $q_{y'}^{\text{max}}/q_{x'}^{\text{max}} \approx 1.21$, where the fixed momentum choices are illustrated by the two cross-section curves in black and red dashed lines in Fig.~\ref{fig:dielectric function}(c) and the corresponding direction-dependent peaks shown in Fig.~\ref{fig:dielectric function}(d). These ratios provide a rough quantitative picture of the anisotropy of the dielectric function, which forbade us from simplifying the dielectric function to a $\norm{\mathbf{q}}$-dependent function. Despite that, we can verify that the dielectric function presents a linear dependence with respect to $\mathbf{q}$ for small $\norm{\mathbf{q}}$ values and goes to $1$ at $\mathbf{q}=\mathbf{0}$, in agreement with Eq.~\eqref{eq:eps linear}, as shown by the fit curves (solid gray) in Figs.~\ref{fig:dielectric function}(b) and \ref{fig:dielectric function}(d). Taking the first-order expansion term of the dielectric function, \textit{i.e.}, by the slope of the fit curves in Figs.~\ref{fig:dielectric function}(b) and \ref{fig:dielectric function}(d), we obtain $r_0\approx 208.2$ \AA ~and $r_0\approx 340.6$ \AA ~for the intralayer contribution in the twist angles of $\theta(1,1) = 21.79^\circ$ and $\theta(1,1) = 46.83^\circ$, respectively. For different cross-section directions, one has that the exact fit value for $r_0^{\ell\ell'}$ is highly dependent on $\angle\mathbf{q}$, which is another anisotropy indicator.

\section{Many-body effects: Excitons and optical response}

\subsection{Dielectric screening of the electron-electron interaction}

In this subsection, we follow an approach very analogous to Sec.~\ref{sec:dielectric function formalism}, but now treating the potential of any $p_z$ electron in TBG, placed at site $\mathbf{r}_1$ and layer $\ell_1$, as an ``external'' potential acting on another $p_z$ electron placed at site $\mathbf{r}_2$ and layer $\ell_2$. We denote this external potential by $\phi_{0}^{\ell_1\ell_2}(\mathbf{r}_1,\mathbf{r}_2)$ and the screened total potential by $\phi^{\ell_1\ell_2}(\mathbf{r}_1,\mathbf{r}_2)$. 

The total potential can be expanded in plane waves applying the Fourier transform (\ref{eq:Fourier transform real space}) with respect to the real space of $\mathbf{r}_2$ positions:
\begin{equation}\label{eq:total potential for external point charge}
    \phi^{\ell_1\ell_2}(\mathbf{r}_1,\mathbf{r}_2) = \dfrac{1}{\mathcal{S}} \sum_{\mathbf{q}}^{\text{BZ}}\sum_{\mathbf{G}} \phi^{\ell_1\ell_2}(\mathbf{r}_1,\mathbf{q}+\mathbf{G}) e^{\mathrm{i} (\mathbf{q}+\mathbf{G}) \cdot \mathbf{r}_2}.
\end{equation}
Rewriting $\phi^{\ell_1\ell_2}(\mathbf{r}_1,\mathbf{q}+\mathbf{G})$ in terms of the dielectric function, according to Eq.~\eqref{eq:dielectric function final}, one has
\begin{equation}\label{eq:Coulomb potential dielectric function inverse}
    \phi^{\ell_1\ell_2}(\mathbf{r}_1,\mathbf{q}+\mathbf{G}) = \sum_{\ell'\mathbf{G}'} {\epsilon^{-1}}^{\ell_2\ell'}_{\mathbf{G}\mathbf{G}'}(\mathbf{q}) \phi_0^{\ell_1\ell'}(\mathbf{r}_1,\mathbf{q}+\mathbf{G}').
\end{equation}
The bare Coulomb potential satisfies $\phi_0^{\ell_1\ell_2}(\mathbf{r}_1,\mathbf{r}_2)=\phi_0^{\ell_1\ell_2}(\mathbf{r}_2-\mathbf{r}_1)$, which allow us to apply the translation property of the Fourier transform
\begin{equation}\label{eq:Fourier transform translation}
    \phi_0^{\ell_1\ell_2}(\mathbf{r}_1,\mathbf{q}+\mathbf{G}) = \phi_0^{\ell_1\ell_2}(|\mathbf{q}+\mathbf{G}|) e^{-\mathrm{i}|\mathbf{q}+\mathbf{G}|\cdot \mathbf{r}_1},
\end{equation}
recognizing that the bare Coulomb potential in momentum space is $\phi^{\ell_1\ell_2}_0(|\mathbf{q}+\mathbf{G}|)=-eX^{\ell_1\ell_2}(|\mathbf{q}+\mathbf{G}|)$ [See Eq.~\eqref{eq:coulomb potential momentum space}]. Replacing Eqs.~\eqref{eq:Coulomb potential dielectric function inverse} and (\ref{eq:Fourier transform translation}) into Eq.~\eqref{eq:total potential for external point charge}, it results in
\begin{equation}
    \phi^{\ell_1\ell_2}(\mathbf{r}_1,\mathbf{r}_2) = -\dfrac{e}{\mathcal{S}} \sum_{\mathbf{q}}^{\text{BZ}} \sum_{\mathbf{G}\mathbf{G}'} \psi^{\ell_1\ell}_{\mathbf{G}\mathbf{G}'}(\mathbf{q}) e^{\mathrm{i} (\mathbf{q}+\mathbf{G})\cdot\mathbf{r}_2} e^{-\mathrm{i}(\mathbf{q}+\mathbf{G}')\cdot\mathbf{r}_1},
\end{equation}
with the definition for the screening term $\psi^{\ell_1\ell_2}_{\mathbf{G}\mathbf{G}'}(\mathbf{q})$ as
\begin{equation}\label{eq:psi term}
    \psi^{\ell_1\ell_2}_{\mathbf{G}\mathbf{G}'}(\mathbf{q}) = \sum_{\ell'} {\epsilon^{-1}}^{\ell_2\ell'}_{\mathbf{G}\mathbf{G}'}(\mathbf{q}) X^{\ell_1\ell'}(|\mathbf{q}+\mathbf{G}'|).
\end{equation}
Then, the interaction between $p_z$ electrons originating from this potential is a two-body operator, written by
\begin{widetext}
\begin{align}\label{eq:Hee}
    \mathcal{H}_{\text{ee}} &= -\dfrac{e}{2} \sum_{\ell_1\ell_2} \sum_{\boldsymbol{\delta}_{\ell_1}\boldsymbol{\delta}_{\ell_2}} \sum_{\mathbf{R}_1\mathbf{R}_2} \phi^{\ell_1\ell_2}(\mathbf{R}_1+\boldsymbol{\delta}_{\ell_1},\mathbf{R}_2+\boldsymbol{\delta}_{\ell_2}) 
    c^\dagger_{\mathbf{R}_1+\boldsymbol{\delta}_{\ell_1}}
    c^\dagger_{\mathbf{R}_2+\boldsymbol{\delta}_{\ell_2}}
    c_{\mathbf{R}_2+\boldsymbol{\delta}_{\ell_2}}
    c_{\mathbf{R}_1+\boldsymbol{\delta}_{\ell_1}}\nonumber \\
    &= \sum_{n_1n_2n_3n_4} \sum_{\mathbf{k}_1\mathbf{k}_2\mathbf{k}_3\mathbf{k}_4}^{\text{BZ}} V^{\mathbf{k}_1\mathbf{k}_2\mathbf{k}_3\mathbf{k}_4}_{n_1n_2n_3n_4} a^\dagger_{n_1\mathbf{k}_1} a^\dagger_{n_2\mathbf{k}_2} a_{n_3\mathbf{k}_3} a_{n_4\mathbf{k}_4},
\end{align}
where the factor of $1/2$ is included to avoid double counting, and we defined the auxiliar term
\begin{equation}
        V_{n_1n_2n_3n_4}^{\mathbf{k}_1\mathbf{k}_2\mathbf{k}_3\mathbf{k}_4} = \dfrac{e^2}{2\mathcal{S}} \sum_{\ell_1\ell_2} \sum_{\mathbf{G}\mathbf{G}'} \psi^{\ell_1\ell_2}_{\mathbf{G}\mathbf{G}'}(\mathbf{k}_1-\mathbf{k}_4) (M_{\ell_1}^{n_1n_4}(\mathbf{k}_4,\mathbf{k}_1-\mathbf{k}_4,\mathbf{G}'))^* M_{\ell_2}^{n_2 n_3}(\mathbf{k}_2,\mathbf{k}_3 - \mathbf{k}_2, \mathbf{G}) \delta_{\mathbf{k}_1-\mathbf{k}_4,\mathbf{k}_3-\mathbf{k}_2}.
\end{equation}
\end{widetext}

\subsection{Semiconductor Bloch Equations}

For this section, we will use a compact notation to soften the burden of the algebraic manipulations and keep the equations reasonably short.
First, we will use the compound index $i\equiv\{n_i,\mathbf{k}_i\}$.
For even more simplicity, we will write the annihilation and creation operators through the simple notation
\begin{equation}
    i \equiv a_{n_i \mathbf{k}_i}.
\end{equation}
In addition, we will also consider implicit summations, in a similar fashion with Einstein's notation.

First we define the full Hamiltonian, written as the sum of three terms:
\begin{equation}\label{eq:H terms}
    \mathcal{H} = \mathcal{H}_0 + \mathcal{H}_I + \mathcal{H}_{\text{ee}}.
\end{equation}
The dipole energy term $\mathcal{H}_I$ is related to the interaction of $p_z$ electrons with the classical electric field $\boldsymbol{\mathcal{E}}$ of the incident light:
\begin{equation} \label{eq:HI compact}
    \mathcal{H}_I =  d_{34} 3^\dagger 4,
\end{equation}
where $d_{34} = \boldsymbol{\mathcal{E}}\cdot \mathbf{d}_{34}$, $\mathbf{d}_{34}$ is the dipole matrix element
\begin{equation}\label{eq:dipole compact}
    \mathbf{d}_{34} = -e \langle 3 | \hat{\mathbf{r}} | 4 \rangle,
\end{equation}
and $\hat{\mathbf{r}}$ is the position operator.
Substituting Eq.~\eqref{eq:HI compact} on Eq.~\eqref{eq:H terms} and rewriting the single-particle [Eq.~\eqref{eq:H0 diagonal}] and electron-electron [Eq.~\eqref{eq:Hee}] Hamiltonians in the compact notation, we obtain
\begin{equation}\label{eq:H}
    \mathcal{H} = E_3 3^\dagger 3 + d_{34}3^\dagger 4 + V_{3456} 3^\dagger 4^\dagger 5 6.
\end{equation}

Next, we introduce the Heisenberg equation of motion
\begin{equation}\label{eq:heisenberg}
    -\mathrm{i}\hbar\partial_t p_{12} = \langle[\mathcal{H},1^\dagger 2]\rangle,
\end{equation}
where $p_{12} = \langle 1^\dagger 2 \rangle$.
Substituting Eq.~\eqref{eq:H} into Eq.~\eqref{eq:heisenberg}, we get
\begin{eqnarray}\label{eq:commutators}
    -i\hbar \partial_t p_{12} = 
    \langle[E_3 3^\dagger 3,1^\dagger 2]\rangle + \langle[d_{34}3^\dagger 4,1^\dagger 2]\rangle + \nonumber\\
    + \langle[V_{3456}3^\dagger 4^\dagger 5 6,1^\dagger 2]\rangle.
\end{eqnarray}
To compute the commutators on the right-hand side, we must apply the anticommutation rules of fermionic creation and annihilation operators:
\begin{subequations} \label{eq:anticomm}
    \begin{equation}
        \{1,2\} = 0, \label{eq:fermion anticommutation rule 1}
    \end{equation}
    \begin{equation}
        \{1^\dagger, 2\} = \delta_{12}. \label{eq:fermion anticommutation rule 2}
    \end{equation}
\end{subequations}
Due to the relation \eqref{eq:fermion anticommutation rule 1}, we also have the property
\begin{equation}\label{eq:V property}
    V_{1234} = V_{2143}
\end{equation}
and the approximation
\begin{equation}\label{eq:RPA}
    \langle 1^\dagger 2^\dagger 3 4 \rangle \approx p_{14}p_{23} - p_{13}p_{24},
\end{equation}
that truncates the equation of motion by neglecting three-particle and higher order terms.
Using Eqs. (\ref{eq:fermion anticommutation rule 2}, \ref{eq:V property}, \ref{eq:RPA}) to simplify Eq.~\eqref{eq:commutators}, it follows that
\begin{eqnarray}\label{eq:SBE general}
    -i\hbar\partial_t p_{12} = (E_1 - E_2) p_{12} + d_{31}p_{32}-d_{23}p_{13} + \nonumber\\
    + 2p_{45}\left[(V_{3451}-V_{3415})p_{32}+(V_{4253}-V_{2453})p_{13}\right].
\end{eqnarray}

Eq.~\eqref{eq:SBE general} represents a general form of the SBE, a set of equations whose solutions are the expectation values of the density matrix elements $p_{12}$.
In the scope of this work, we are solely concerned with transitions between the uppermost valence ($v$) and the lowermost conduction band ($c$).
This is commonly referred to as the two-band approximation.
Moreover, momentum is conserved in the electronic transitions since the electric field is homogeneous over all the sample.
This allows us to write
\begin{equation}\label{eq:momentum conservation}
    p_{12} \equiv p_{n_1n_2}^{\mathbf{k}_1\mathbf{k}_2} = p_{n_1n_2}^{\mathbf{k}_1\mathbf{k}_1} \delta_{\mathbf{k}_1\mathbf{k}_2}.
\end{equation}
Now we rewrite Eq.~\eqref{eq:SBE general} going back to the previous notation and applying Eq.~\eqref{eq:momentum conservation}:
\begin{eqnarray}\label{eq:SBE expanded notation}
    -i\hbar \partial_t p_{n_1n_2}^{\mathbf{k}\mathbf{k}}=
    (E_{n_1\mathbf{k}} - E_{n_2\mathbf{k}})p_{n_1n_2}^{\mathbf{k}\mathbf{k}} + \nonumber\\
    + \sum_{n_3} d_{n_3n_1}^{\mathbf{k}\mathbf{k}}p_{n_3n_2}^{\mathbf{k}\mathbf{k}} - d_{n_2n_3}^{\mathbf{k}\mathbf{k}} p_{n_1n_3}^{\mathbf{k}\mathbf{k}} + \nonumber\\
    + 2 \sum_{n_3n_4n_5} \sum_{\mathbf{k}_4}^{\text{BZ}} p_{n_4n_5}^{\mathbf{k}_4\mathbf{k}_4} \times \nonumber \\ 
    \left[ (V_{n_3n_4n_5n_1}^{\mathbf{k}\mathbf{k}_4\mathbf{k}_4\mathbf{k}} - V_{n_3n_4n_1n_5}^{\mathbf{k}\mathbf{k}_4\mathbf{k}\mathbf{k}_4})p_{n_3n_2}^{\mathbf{k}\mathbf{k}} + \right. \nonumber\\
    \left. + (V_{n_4n_2n_5n_3}^{\mathbf{k}_4\mathbf{k}\mathbf{k}_4\mathbf{k}} - V_{n_2n_4n_5n_3}^{\mathbf{k}\mathbf{k}_4\mathbf{k}_4\mathbf{k}}) p_{n_1n_3}^{\mathbf{k}\mathbf{k}} \right],
\end{eqnarray}
where the summations $\sum_{n_i}$ are restricted to $n_i=c,v$.
For the purposes of this work, it will be enough to obtain the solution of the SBE for the matrix element $p_{cv}^{\mathbf{k}\mathbf{k}}$.
For the dipole energy terms, intraband transitions, \textit{i.e.}, terms with factors of $d_{cc}^{\mathbf{k}\mathbf{k}}$ or $d_{vv}^{\mathbf{k}\mathbf{k}}$, can be safely neglected.
Moreover, terms involving $p_{vc}^{\mathbf{k}\mathbf{k}}$ will be neglected, in light of the Rotating Wave Approximation (RWA).
We also neglect nonlinear terms, \textit{i.e.}, terms where factors of $(p_{n_i n_j}^{\mathbf{k}\mathbf{k}})^2$ appear.
In the end, we remember that the condition of charge neutrality implies $p_{vv}^{\mathbf{k}\mathbf{k}} = 1$ and $p_{cc}^{\mathbf{k}\mathbf{k}} = 0$.
Under all these considerations, Eq.~\eqref{eq:SBE expanded notation} simplifies to
\begin{eqnarray}\label{eq:SBE explicit}
    -i\hbar\partial_t p_{cv}^{\mathbf{k}\mathbf{k}} = (E_{c\mathbf{k}} - E_{v\mathbf{k}}) p_{cv}^{\mathbf{k}\mathbf{k}} + d_{vc}^{\mathbf{k}\mathbf{k}} + 2\sum_{\mathbf{k}_4}^{\text{BZ}} \left[p_{cv}^{\mathbf{k}\mathbf{k}} \times \right. \nonumber\\
    \left. \times (V_{cvvc}^{\mathbf{k}\mathbf{k}_4\mathbf{k}_4\mathbf{k}} - V_{cvcv}^{\mathbf{k}\mathbf{k}_4\mathbf{k}\mathbf{k}_4} + V_{vvvv}^{\mathbf{k}_4\mathbf{k}\mathbf{k}_4\mathbf{k}} - V_{vvvv}^{\mathbf{k}\mathbf{k}_4\mathbf{k}_4\mathbf{k}}) + \right. \nonumber\\
    \left. + p_{cv}^{\mathbf{k}_4}(V_{vcvc}^{\mathbf{k} \mathbf{k}_4 \mathbf{k}_4 \mathbf{k}} - V_{vccv}^{\mathbf{k}\mathbf{k}_4\mathbf{k}\mathbf{k}_4})\right].
\end{eqnarray}
Now we recognize the terms
\begin{subequations}\label{eq:SBE terms}
    \begin{align}\label{eq:exchange self-energy}
        \Sigma_{\mathbf{k}} \equiv 2\sum_{\mathbf{k}_4}^{\text{BZ}} V_{cvvc}^{\mathbf{k}\mathbf{k}_4\mathbf{k}_4\mathbf{k}} - V_{cvcv}^{\mathbf{k}\mathbf{k}_4\mathbf{k}\mathbf{k}_4} + V_{vvvv}^{\mathbf{k}_4\mathbf{k}\mathbf{k}_4\mathbf{k}} - V_{vvvv}^{\mathbf{k}\mathbf{k}_4\mathbf{k}_4\mathbf{k}},
    \end{align}
    \begin{equation}\label{eq:kernel}
        K(\mathbf{k},\mathbf{k}_4) \equiv -2\mathcal{S}(V_{vcvc}^{\mathbf{k} \mathbf{k}_4 \mathbf{k}_4 \mathbf{k}} - V_{vccv}^{\mathbf{k}\mathbf{k}_4\mathbf{k}\mathbf{k}_4}).
    \end{equation}
\end{subequations}

The term $\Sigma_{\mathbf{k}}$ physically represents the exchange self-energy correction, which corrects the optical band of the system.
Using $\Sigma_{\mathbf{k}}$ we write the renormalized optical band as
\begin{equation}\label{eq:optical band}
    \hbar\tilde \omega_{\mathbf{k}} = E_{c\mathbf{k}} - E_{v\mathbf{k}} + \Sigma_{\mathbf{k}}.
\end{equation}

The term $K(\mathbf{k},\mathbf{k}_4)$, on the other hand, is commonly referred as the kernel of the SBE.
Using these definitions, explicitating the dipole energy term $d_{vc}^{\mathbf{k}\mathbf{k}} = \boldsymbol{\mathcal{E}}\cdot \mathbf{d}_{vc}^{\mathbf{k}\mathbf{k}}$, and replacing $\partial_t \rightarrow i\omega$, where $\omega$ is the frequency of oscillation of the electric field, we can rewrite Eq.~\eqref{eq:SBE explicit} as, finally, the final form of our SBE:
\begin{equation}\label{eq:SBE final}
    \hbar(\omega-\tilde\omega_{\mathbf{k}}) p_{cv}(\mathbf{k}) + \dfrac{1}{\mathcal{S}}\sum_{\mathbf{q}}^{\text{BZ}} K(\mathbf{k},\mathbf{q}) p_{cv}(\mathbf{q}) = \boldsymbol{\mathcal{E}} \cdot \mathbf{d}_{vc}(\mathbf{k}),
\end{equation}
where $p_{cv}(\mathbf{k}) \equiv p_{cv}^{\mathbf{k}\mathbf{k}}$ and $\mathbf{d}_{vc}(\mathbf{k}) \equiv \mathbf{d}_{vc}^{\mathbf{k}\mathbf{k}}$.
This is a linear integral equation for $p_{cv}(\mathbf{k})$.
Setting the independent term on the right-hand side to zero is equivalent to solving the equation of motion without the dipole energy term, whose solutions will give us the exciton states that the material can host.
In this case, $\hbar\omega$ will become the exciton eigenenergies, and $p_{cv}(\mathbf{k})$, the exciton wavefunctions, forming an eigenvalue problem.
When the dipole energy term is included, however, we will investigate the optical response of the material for incidence of monocromatic light with frequency $\omega$, and $p_{cv}(\mathbf{k})$ becomes the interband transition amplitude.
Performing the replacement $\hbar\omega\rightarrow\hbar\omega+i\gamma$, were $\gamma$ is a phenomenological term for the relaxation transition rate, and rewriting the summation in $\mathbf{q}$ in the continuum limit $\sum_{\mathbf{q}} \rightarrow \mathcal{S} \int \text{d}^2\mathbf{q}/(2\pi)^2 $, we rewrite the SBE in the form that appears on the main text:
\begin{equation}\label{eq:SBE continuum}
    (\hbar\omega-\hbar\tilde\omega_{\mathbf{k}} + i\gamma) p_{cv}(\mathbf{k}) + \int \dfrac{\text{d}^2\mathbf{q}}{(2\pi)^2} K(\mathbf{k},\mathbf{q}) p_{cv}(\mathbf{q}) = \boldsymbol{\mathcal{E}} \cdot \mathbf{d}_{vc}(\mathbf{k}).
\end{equation}

As a last discussion of this subsection, we will simplify the terms \eqref{eq:SBE terms} even further.
First, we notice that terms with factors of the type $V_{n_3n_4n_5n_6}^{\mathbf{k}\mathbf{k}_4\mathbf{k}_4\mathbf{k}}$ can be safely neglected taking into account that, for the sums in $\mathbf{G},\mathbf{G}'$, the terms $\mathbf{G}=\mathbf{G}'=\mathbf{0}$ will be exactly compensated by the charged background of ions, and the remaining terms will not have any $\mathbf{q}$-dependence but will decay as $1/|\mathbf{G}|$ and can be safely neglected.
This allow us to write the exchange self-energy and the kernel as
\begin{widetext}
\begin{subequations}
    \begin{align}
        \Sigma_{\mathbf{k}} &= 2\sum_{\mathbf{q}}^{\text{BZ}} V_{vvvv}^{\mathbf{k}+\mathbf{q},\mathbf{k},\mathbf{k}+\mathbf{q},\mathbf{k}} - V_{vcvc}^{\mathbf{k}+\mathbf{q},\mathbf{k},\mathbf{k}+\mathbf{q},\mathbf{k}}\\
        &= e^2 \int \dfrac{\text{d}^2\mathbf{q}}{(2\pi)^2} \sum_{\ell_1\ell_2}\sum_{\mathbf{G}\mathbf{G}'} \psi^{\ell_1\ell_2}_{\mathbf{G}\mathbf{G}'}(\mathbf{q}) \times 
        \left[ (M_{\ell_1}^{vv}(\mathbf{k},\mathbf{q},\mathbf{G}'))^* M_{\ell_2}^{vv}(\mathbf{k},\mathbf{q},\mathbf{G}) - (M_{\ell_1}^{vc}(\mathbf{k},\mathbf{q},\mathbf{G}'))^* M_{\ell_2}^{cv}(\mathbf{k},\mathbf{q},\mathbf{G}) \right],
    \end{align}
    \begin{align}
    K(\mathbf{k},\mathbf{q}) &= 2\mathcal{S}V_{vccv}^{\mathbf{k}\mathbf{q}\mathbf{k}\mathbf{q}} = 2\mathcal{S}V_{cvvc}^{\mathbf{q}\mathbf{k}\mathbf{q}\mathbf{k}}\\
        &= e^2\sum_{\ell_1\ell_2} \sum_{\mathbf{G}\mathbf{G}'} \psi^{\ell_1\ell_2}_{\mathbf{G}\mathbf{G}'}(\mathbf{q}-\mathbf{k}) (M_{\ell_1}^{cc}(\mathbf{k},\mathbf{q}-\mathbf{k},\mathbf{G}'))^*  M_{\ell_2}^{vv}(\mathbf{k},\mathbf{q}-\mathbf{k},\mathbf{G}).
    \end{align}
\end{subequations}
\end{widetext}

Second, we will expand the dipole moment. For this, we note that
\begin{equation}
\hat{\mathbf{r}}|\mathbf{R}+\boldsymbol{\delta}_\ell\rangle=(\mathbf{R}+\boldsymbol{\delta}_\ell)|\mathbf{R}+\boldsymbol{\delta}_\ell\rangle,
\end{equation}
and so:
\begin{equation}
\hat{\mathbf{r}}|n\mathbf{k}\rangle= \sum_{\mathbf{R},\boldsymbol{\delta}_\ell}
(\mathbf{R}+\boldsymbol{\delta}_\ell)
e^{i(\mathbf{R}+\boldsymbol{\delta}_\ell)\cdot\mathbf{k}}u_{n\boldsymbol{\delta}_\ell}(\mathbf{k})
|\mathbf{R}+\boldsymbol{\delta}_\ell\rangle.
\end{equation}
Defining
\begin{equation}
\mathcal{H}_\mathbf{k}=\sum_n E_{n\mathbf{k}} |n\mathbf{k}\rangle\langle n\mathbf{k}|,
\end{equation}
we have
\begin{eqnarray}
[\mathcal{H}_\mathbf{k},\hat{\mathbf{r}}]=\sum_n E_{n\mathbf{k}} 
\sum_{\mathbf{R},\boldsymbol{\delta}_\ell}\left[ 
(\mathbf{R}+\boldsymbol{\delta}_\ell)
-
(\mathbf{R}^\prime+\boldsymbol{\delta}^\prime_\ell)\right]
\nonumber\\ 
e^{i(\mathbf{R}+\boldsymbol{\delta}_\ell)\cdot\mathbf{k}}u_{n\boldsymbol{\delta}_\ell}(\mathbf{k})e^{-i(\mathbf{R}^\prime+\boldsymbol{\delta}^\prime_\ell)\cdot\mathbf{k}}u^*_{n{\boldsymbol{\delta}_\ell}^\prime}(\mathbf{k})
\nonumber\\
|\mathbf{R}+\boldsymbol{\delta}_\ell\rangle\langle \mathbf{R}^\prime+\boldsymbol{\delta}^\prime_{\ell^\prime}|,
\end{eqnarray}
that can be rewritten as:
\begin{eqnarray}
[\mathcal{H}_\mathbf{k},\hat{\mathbf{r}}]=\nabla_\mathbf{k}\mathcal{H}_\mathbf{k}
-\sum_n\left[ \nabla_\mathbf{k}\left( E_{n\mathbf{k}}\right)|\mathbf{k}n\rangle\langle \mathbf{k}n|\right.
-\nonumber\\ \left.
-\sum_{\mathbf{R},\boldsymbol{\delta}_\ell} \sum_{\mathbf{R}^\prime,\boldsymbol{\delta}_\ell^\prime} 
\nabla_\mathbf{k}\left(
u^*_{n{\boldsymbol{\delta}_\ell}^\prime}(\mathbf{k})
u_{n\boldsymbol{\delta}_\ell}(\mathbf{k}) \right)\right.
\nonumber\\   \left.
e^{-i(\mathbf{R}^\prime+\boldsymbol{\delta}^\prime_\ell)\cdot\mathbf{k}}e^{i(\mathbf{R}+\boldsymbol{\delta}_\ell)\cdot\mathbf{k}}  
|\mathbf{R}+\boldsymbol{\delta}_\ell\rangle\langle \mathbf{R}^\prime+\boldsymbol{\delta}^\prime_{\ell^\prime}|\right].
\end{eqnarray}

Now, for $m\neq n$,
\begin{flalign}
\langle m\mathbf{k}|[\mathcal{H}_\mathbf{k},\hat{\mathbf{r}}]|n\mathbf{k}\rangle=\langle m\mathbf{k}|\nabla_\mathbf{k}\mathcal{H}_\mathbf{k}|n\mathbf{k}\rangle
-\nonumber\\ 
\sum_{n^\prime}\sum_{\mathbf{R},\boldsymbol{\delta}_\ell} \sum_{\mathbf{R}^\prime,\boldsymbol{\delta}_\ell^\prime} 
\nabla_\mathbf{k}\left(
u^*_{n^\prime{\boldsymbol{\delta}_\ell}^\prime}(\mathbf{k})
u_{n^\prime\boldsymbol{\delta}_\ell}(\mathbf{k}) \right)
u^*_{m\boldsymbol{\delta}_\ell}(\mathbf{k})u_{n\boldsymbol{\delta}_\ell^\prime},
\end{flalign}
and we can show that
\begin{equation}
\sum_{n^\prime}\sum_{\mathbf{R},\boldsymbol{\delta}_\ell} \sum_{\mathbf{R}^\prime,\boldsymbol{\delta}_\ell^\prime} 
\nabla_\mathbf{k}\left(
u^*_{n^\prime{\boldsymbol{\delta}_\ell}^\prime}(\mathbf{k})
u_{n^\prime\boldsymbol{\delta}_\ell}(\mathbf{k}) \right)
u^*_{m\boldsymbol{\delta}_\ell}(\mathbf{k})u_{n\boldsymbol{\delta}_\ell^\prime}=0.
\end{equation}

Thus,
\begin{equation}
\langle m\mathbf{k}|[\mathcal{H}_\mathbf{k},\hat{\mathbf{r}}]|n\mathbf{k}\rangle=\langle m\mathbf{k}|\nabla_\mathbf{k} \hat{H}_\mathbf{k}|n\mathbf{k}\rangle,
\end{equation}
and using that
\begin{equation}
\langle n \mathbf{k}| [\mathcal{H}_{\mathbf{k}},\hat{\mathbf{r}}] |m \mathbf{k}\rangle=
(E_{n\mathbf{k}}-E_{m\mathbf{k}})\langle n \mathbf{k}|\hat{\mathbf{r}} |m \mathbf{k}\rangle,
\end{equation}
we obtain, finally,
\begin{equation}
\mathbf{d}_{vc}(\mathbf{k})=e\frac{\langle v\mathbf{k} |\nabla_\mathbf{k} \mathcal{H}_\mathbf{k}| c\mathbf{k}\rangle}
{E_{c\mathbf{k}}-E_{v\mathbf{k}}}. \label{eq:dipole moment}
\end{equation}

\subsection{Optical band, binding energies, exciton wavefunctions, and its layer hybridization}

\begin{figure}[t]
    \centering
    \includegraphics[width=\linewidth]{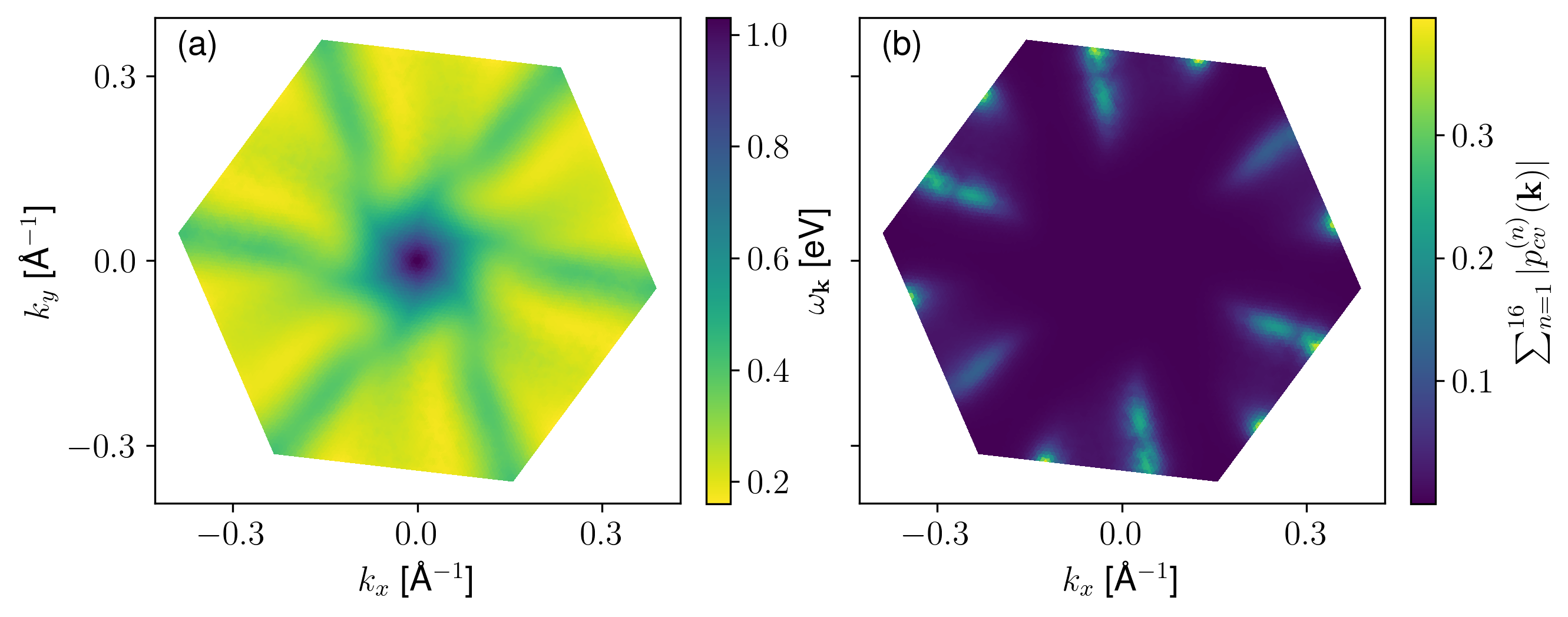}
    \caption{
        (Color online) Renormalized optical band structure [Eq.~\eqref{eq:optical band}] of the TBG for (a) $\theta(1,6) = 46.83^\circ$, $V=3$ eV and $d=2.8$ \AA, exhibiting a six-fold shape, resembling a flower, with twelve degenerate minima located at the edges of the Brillouin zone of the supercell, and (b) a merging plot of the sixteen first excitonic wavefunctions in momentum space, \textit{i.e.} $\sum_{n=1}^{16} |p_{cv,0}^{(0)}(\mathbf{k})|$, obtained by solving the homogeneous SBE, also known as the Bethe-Salpeter equation [Eq.~\eqref{eq:SBE final}]. The excitonic wavefunctions present stretched dome shapes located precisely in the optical band valleys, \textit{i.e.} the wavefunction peaks positioned in momentum space in the minima of the ``leaves'' of optical band edges.}
    \label{fig:optical band}
\end{figure}

By numerically computing Eq.~\eqref{eq:optical band}, we obtained the renormalized optical bands of the TBG, defined as the transition energy between the lowest conduction and highest valence bands with a G$_0$W$_0$ gap correction, for $\theta(1,1)=21.79^\circ$, $V=3$ eV, and $2.6$ \AA ~ shown in Fig.~\textcolor{blue}{2(a)} of the main text and for $\theta(1,6) = 46.83^\circ$, $V=3$ eV and $d=2.8$ \AA ~in Fig.~\ref{fig:optical band}(a). In both twist angle cases, one observes that the TBG optical bands' profiles exhibit interesting six-fold symmetric patterns with a maximum localized in the $\mathbf{\Gamma}$--point. In addition to what is discussed in the main text for the $\theta(1,1)=21.79^\circ$--case [Fig.~\textcolor{blue}{2(a)}], one notices for $\theta(1,6) = 46.83^\circ$ [Fig.~\ref{fig:optical band}(a)] that the optical band resembles a flower of six petals with twelve degenerate minima located at the edges of the Brillouin zone of the supercell. By evaluating numerically the self-energy values $\sum_{\mathbf{k}}$, we observed that such gap correction by itself does not change the overall qualitative behavior of the optical bands. However, it dramatically increases the energy gap, corroborating that band gap renormalizations are very significant for calculating quasiparticle excitations in 2D materials and must be considered to make accurate predictions.

\begin{figure}[!t]
    \centering
    \includegraphics[width=0.49\linewidth]{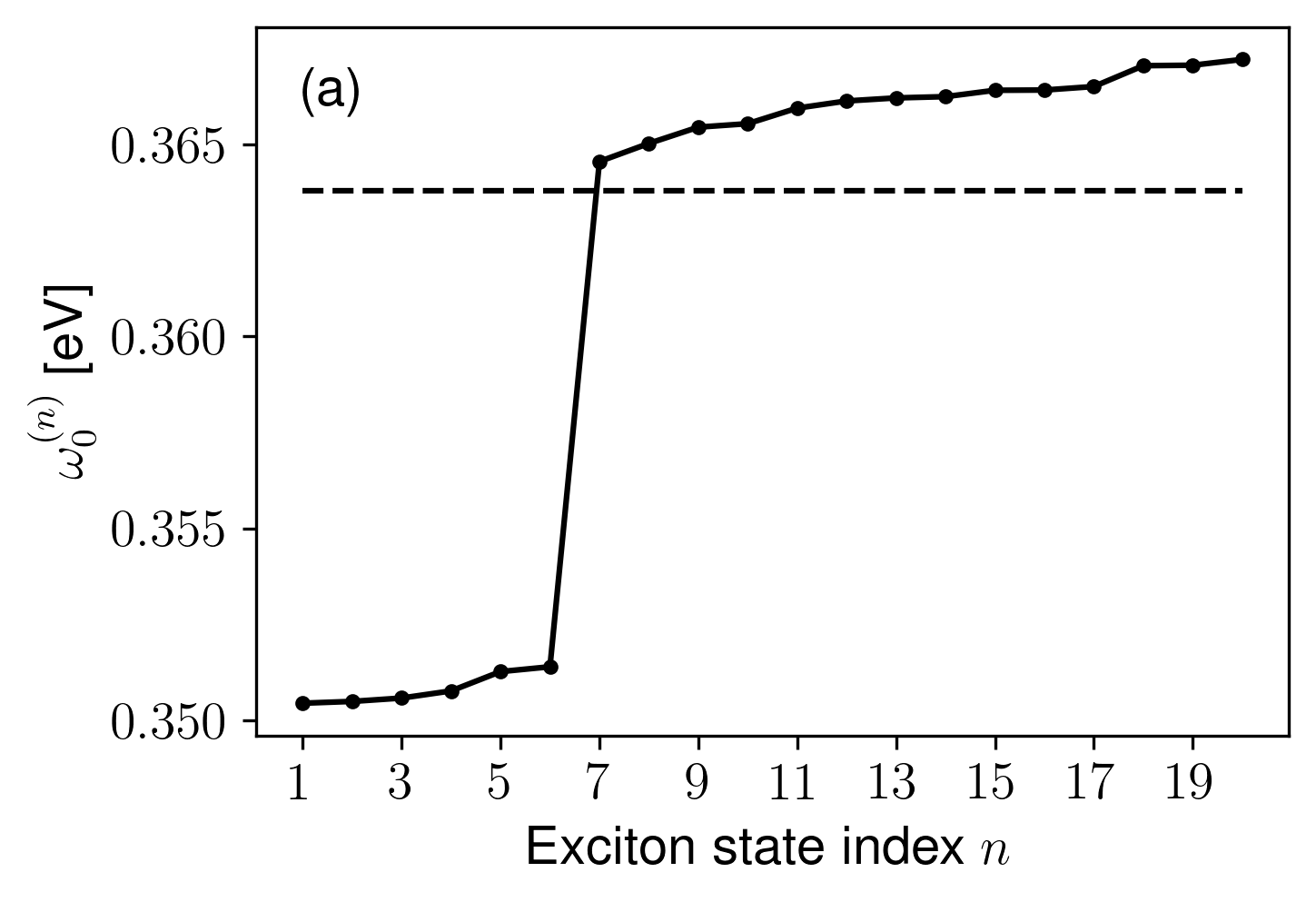}
    \includegraphics[width=0.49\linewidth]{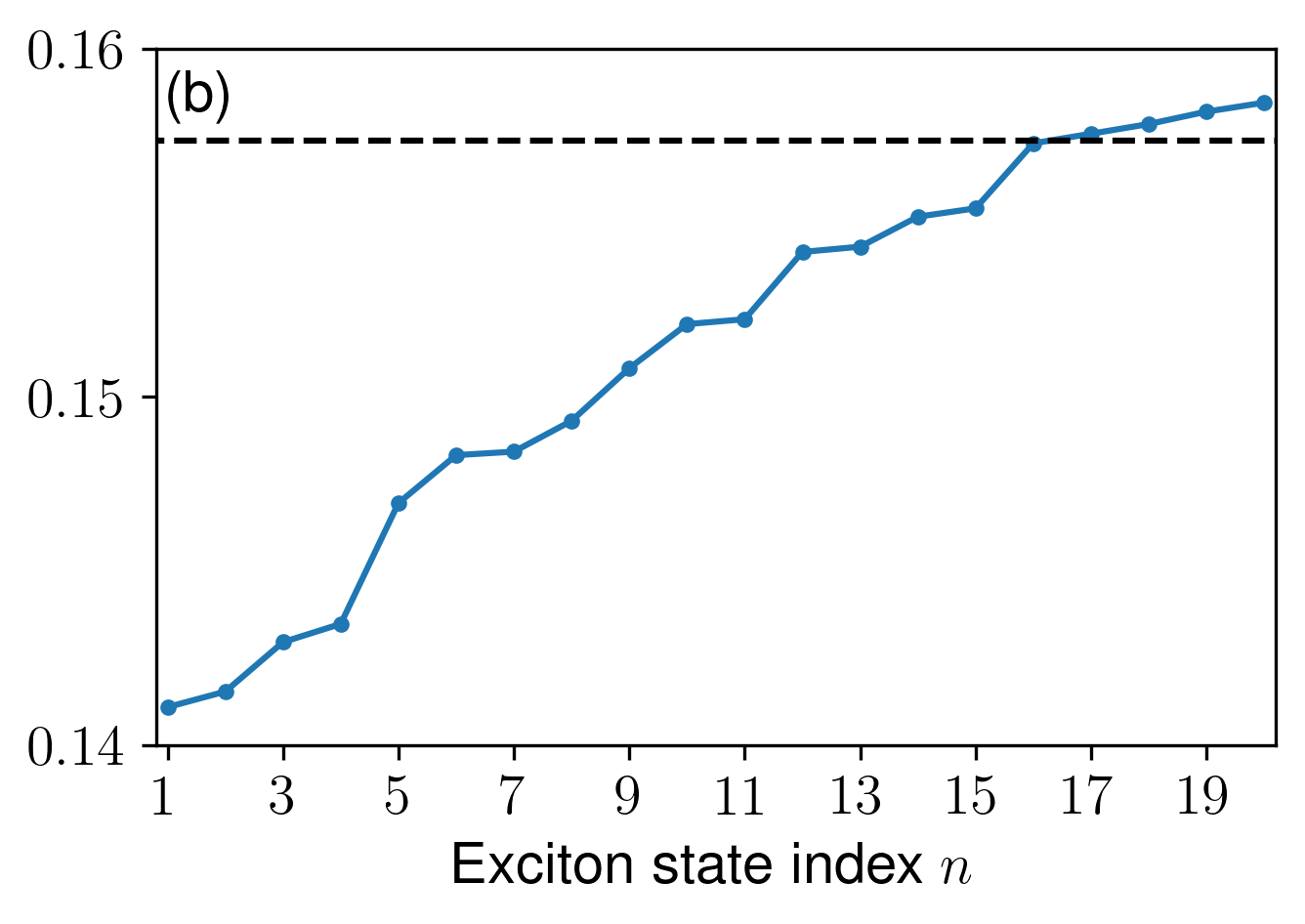}
    \caption{(Color online) The first 20th excitonic energies of TBG, taking $N_h=120$ (see Eq.~\eqref{eq:k-point sampling hexagonal}), for (a) $\theta(1,1)=21.79^\circ$, $V=3$ eV, and $d = 2.6$ \AA, ~and (b) $\theta(1,6) = 46.83^\circ$, $V=3$ eV and $d=2.8$ \AA. The dashed lines indicate the optical band minima in each case. A clear six-fold degeneracy is observed in panel (a) in agreement with the $C_6$ system symmetry.}
    \label{fig:spectrum}
\end{figure}

The exciton wavefunctions $p_{cv,0}^{(n)}(\mathbf{k})$ and energies $\omega_0^{(n)}$ ($n=1,2,3,\hdots$) are obtained by setting the right-hand side of Eq.~\eqref{eq:SBE final} to zero and solving it as an eigenvalue problem. Performing such calculations, we show in Figs.~\ref{fig:spectrum}(a) and \ref{fig:spectrum}(b) the first 20th excitonic energies for (a) $\theta(1,1)=21.79^\circ$, $V=3$ eV, and $d = 2.6$ \AA ~and (b) $\theta(1,6) = 46.83^\circ$, $V=3$ eV and $d=2.8$ \AA. The dashed lines in Figs.~\ref{fig:spectrum}(a) and \ref{fig:spectrum}(b) correspond to the optical band minimum $\min_{\mathbf{k}}\omega_{\mathbf{k}}$ for the assumed sampling $\mathbf{k}$-points. By a convergence analysis regarding the assumed grid points and spacing of the 2D reciprocal space, we verified that the converged values of the optical band minimum are $\min_{\mathbf{k}}\omega_{\mathbf{k}}=364$ meV and $\min_{\mathbf{k}}\omega_{\mathbf{k}}=135$ meV for [Fig.~\ref{fig:spectrum}(a)] $\theta(1,1)=21.79^\circ$ and [Fig.~\ref{fig:spectrum}(b)] $\theta(1,6) = 46.83^\circ$, respectively. From Figs.~\ref{fig:spectrum}(a) and \ref{fig:spectrum}(b), twelve and sixteen excitonic bound states are observed, respectively, \textit{i.e.}, excitons with energies below the optical band minimum. The lower exciton wavefunctions must be localized near the most likely formation points in reciprocal space, \textit{i.e.}, the points where the optical band is minimum. Figures~\ref{fig:optical band}(b) and \textcolor{blue}{2(b)} in the main text present contour plots merging the sixteen $\left[\sum_{n=1}^{16} |p_{cv,0}^{(0)}(\mathbf{k})|\right]$ and twelve $\left[\sum_{n=1}^{12} |p_{cv,0}^{(0)}(\mathbf{k})|\right]$ lowest excitonic wavefunctions, respectively, for $\theta(1,6) = 46.83^\circ$ and $\theta(1,1)=21.79^\circ$ cases, where their amplitudes are maximum at the region in the reciprocal space where the optical bands are minima.

By performing a convergence study of the lowest exciton energy $\omega_0^{(1)}$ for $\theta(1,1)=21.79^\circ$, $V=3$  eV, and $d=2.6$ \AA, as a function of the number of sampling points, one gets that $\omega_0^{(1)}\rightarrow 350.4$ meV, resulting in the converged binding energy of $\min_{\mathbf{k}}\omega_{\mathbf{k}}-\omega_0^{(1)} = 13.3$ meV for the lowest exciton, which corresponds to an intermediate value in the sense that it is much greater than the exciton binding energies found in 3D materials, yet much smaller than the binding energies of 2D semiconductors with greater gaps. This binding energy, however, indicates that TBG can hold excitons with robust binding energies.

\begin{figure}[!h]
    \centering
    \includegraphics[width=\linewidth]{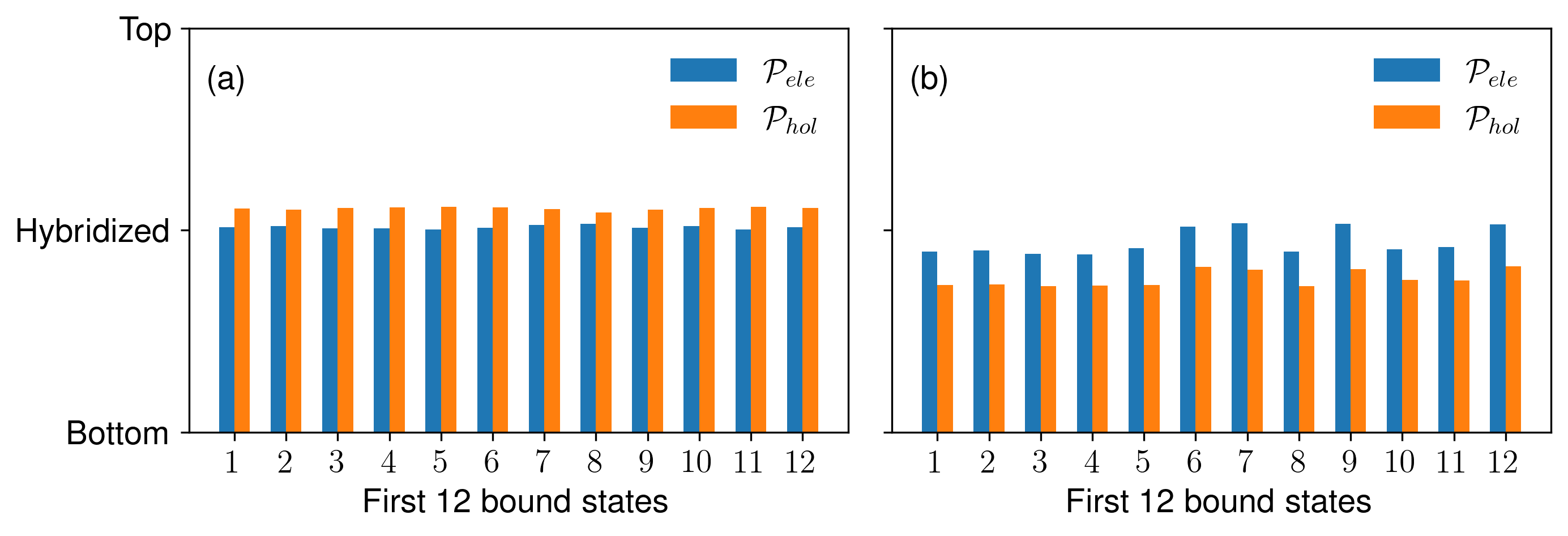}
    \caption{(Color online) Bar plot of electron ($\mathcal{P}_{\text{ele}}$ - blue) and hole ($\mathcal{P}_{\text{hol}}$ - orange) probabilities of being localized in the top or bottom layers of TBG [see Eqs.~\eqref{eq:pele} and \eqref{eq:phol}], for (a) $\theta(1,1)=21.79^\circ$, $V=3$ eV, and $d = 2.6$ \AA ~and (b) $\theta(1,6) = 46.83^\circ$, $V=3$ eV and $d=2.8$ \AA. We used $N_{\mathbf{k}}=8600$ sampling $\mathbf{k}$-points. The probabilities of the twelve first excitonic bound states are close to 0.5, indicating high exciton layer hybridization.}
    \label{fig:hybrid}
\end{figure}

To investigate how the excitons are arranged in TBG, we evaluate the electron and hole layer compositions for the twelve first excitonic bound states. The electron and hole probabilities of each exciton state to be located in the top layer of TBG are given, respectively, by
\begin{subequations}
\begin{align}
    & \mathcal{P}_{\text{ele}}^{(n)} = \sum_{\mathbf{k}}^{\text{BZ}} \sum_{\boldsymbol{\delta}_2} |u_{c\boldsymbol{\delta}_2}(\mathbf{k}) p_{cv,0}^{(n)}(\mathbf{k})|^2, \label{eq:pele} \\
    & \mathcal{P}_{\text{hol}}^{(n)} = \sum_{\mathbf{k}}^{\text{BZ}} \sum_{\boldsymbol{\delta}_2} |u_{v\boldsymbol{\delta}_2}(\mathbf{k}) p_{cv,0}^{(n)}(\mathbf{k})|^2, \label{eq:phol}
\end{align}
\end{subequations}
where $u_{v\boldsymbol{\delta}_{\ell}}(\mathbf{k})$ and $u_{c\boldsymbol{\delta}_{\ell}}(\mathbf{k})$ are the single-particle electron wavefunctions for the highest valence and lowest conduction bands, respectively. Probabilities close to $1$ ($0$) indicate electron or hole localization in the top (bottom) layer. However, intermediate values close to $0.5$ will indicate the hybridization of the electrons or holes between the TBG layers. In Fig.~\ref{fig:hybrid}, we show the exciton hybridization between layers for (a) $\theta(1,1)=21.79^\circ$, $V=3$ eV, and $d = 2.6$ \AA ~and (b) $\theta(1,6) = 46.83^\circ$, $V=3$ eV and $d=2.8$ \AA, evaluated using the probabilities that electrons and holes are localized in one layer or another according to Eq.~\eqref{eq:pele} and \eqref{eq:phol}. As noticed, all excitonic bound states' probabilities are mixed with values close to 0.5, demonstrating that such bound excitons in TBG are strongly layer hybridized, suggesting its interlayer nature, as previously discussed in Figs.~\textcolor{blue}{1(c)} and \textcolor{blue}{1(d)} in the main text.

\subsection{Optical Response}

To analyze the optical response, we calculate the expectation value of the polarization operator
\begin{equation}\label{eq.polarization}
\hat{\mathbf{P}}= -e\sum_{\mathbf{R},\boldsymbol{\delta}_\ell} \hat{\Psi}^\dagger(\mathbf{R},\boldsymbol{\delta}_\ell,t)\left(\mathbf{R}+\boldsymbol{\delta}_\ell\right)\hat{\Psi}^\dagger(\mathbf{R},\boldsymbol{\delta}_\ell,t), 
\end{equation}
with the field operator given by
\begin{equation}\label{eq.field_operator}
\hat{\Psi}(\mathbf{R},\boldsymbol{\delta}_\ell,t)=\frac{1}{\sqrt{N_c}}\sum_n\int \frac{d^2\mathbf{k}}{(2\pi)^2}e^{i\mathbf{k}\cdot\mathbf{R}}u_{n,\boldsymbol{\delta}_\ell}(\mathbf{k}) \hat{a}_{n,\mathbf{k}}.
\end{equation}

Replacing Eq.~\eqref{eq.field_operator} into Eq.~\eqref{eq.polarization}, the polarization operator becomes
\begin{flalign}
\hat{\mathbf{P}}=\frac{e}{N_c}
\sum_{n,n^\prime} \int \frac{d^2\mathbf{k}}{(2\pi)^2} \frac{d^2\mathbf{k}^\prime}{(2\pi)^2}\sum_{\mathbf{R},\boldsymbol{\delta}_\ell} 
e^{-i\mathbf{k}^\prime\cdot\mathbf{R}}u^*_{n^\prime,\boldsymbol{\delta}_\ell}(\mathbf{k}^\prime) \nonumber\\
e^{i\mathbf{k}\cdot\mathbf{R}}u_{n,\boldsymbol{\delta}_\ell}(\mathbf{k}) 
\hat{a}^\dagger_{n^\prime,\mathbf{k}^\prime}
\hat{a}_{n,\mathbf{k}}\left(\mathbf{R}+\boldsymbol{\delta}_\ell\right).
\end{flalign}

Using the dipole matrix element \eqref{eq:dipole moment}, we arrive at the expectation value of the polarization written as:
\begin{equation}
\langle\hat{\mathbf{P}}\rangle= \sum_{n,\mathbf{k}} \sum_{n^\prime,\mathbf{k}^\prime} 
\mathbf{d}_{nn^\prime}^{\mathbf{k}\mathbf{k}^\prime} p_{nn^\prime}^{\mathbf{k}\mathbf{k}^\prime}(t),
\end{equation}
where we defined
\begin{equation}\label{eq.pnk}
p_{nn^\prime}^{\mathbf{k}\mathbf{k}^\prime}(t)\equiv
\langle \hat{a}^\dagger_{\mathbf{k}n}(t) \hat{a}_{\mathbf{k}^\prime n^\prime}(t)  \rangle.
\end{equation}

Using the local approximation \eqref{eq:momentum conservation} and the linear regime on the frequency domain, Eq.~\eqref{eq.pnk} becomes
\begin{equation}
p_{nn^\prime}^{\mathbf{k}\mathbf{k}^\prime}(t)
\approx p_{nn^\prime}^{\mathbf{k}}(\omega)\delta_{\mathbf{k},\mathbf{k}^\prime}e^{i\omega t}.
\end{equation}

Since we seek the optical response near the excitonic energy inside the bandgap, we shall limit ourselves to the inclusion of the uppermost valence and lowest conduction bands, similarly as considered in previous sections, such as
\begin{equation}
\langle\hat{\mathbf{P}}\rangle=\int\frac{d^2\mathbf{k}}{(2\pi)^2}
\mathbf{d}_{cv}^{\mathbf{k}} p_{cv}(\mathbf{k})(\omega)e^{i\omega t},
\end{equation}
now, we decompose the interband transition amplitude into the excitonic basis and the continuum part:
\begin{equation}
p_{cv}^{\mathbf{k}\mathbf{k}^\prime}(\omega)= \sum_j c_j(\omega) \phi_j +\phi_c(\mathbf{k}), \label{eq:decompose}
\end{equation}
where $\phi_j$ is the $j$th solution of the homogeneous version of Eq.~\eqref{eq:SBE final} and $\phi_c$ denotes the continuum part, corresponding to the states above the bandgap, that are orthogonal to any excitonic state $\phi_j$.

Replacing back \eqref{eq:decompose} into Eq.~\eqref{eq:SBE final}, and using the orthogonality property of the exciton wavefunction, we arrive at the following expression for the coefficient $c_j(\omega)$:
\begin{equation}
c_j(\omega)= \frac{\mathbf{d}_j \cdot
\boldsymbol{\mathcal{E}}}
{\hbar\omega-E_j+i\gamma},
\end{equation}
where we introduced a phenomenological relaxation term $\gamma$ to account for nonradiative transitions, and we defined the exciton dipole $d_j$ moment as:
\begin{equation}
\mathbf{d}_j=\int\frac{d^2\mathbf{k}}{(2\pi)^2} \phi^*_j(\mathbf{k}) \mathbf{d}_{cv}(\mathbf{k}).
\end{equation}

The polarization vector becomes:
\begin{equation}\label{eq.P}
\mathbf{P}(\omega)=\sum_j \frac{\mathbf{d}_j \,\mathbf{d}^*_j\cdot\mathbf{\cal E}}{\hbar\omega-E_j+i\gamma}.
\end{equation}

Plugging the classical electromagnetism relations $\mathbf{J}=\partial_t \mathbf{P}$ and $\mathbf{J}=\sigma\mathbf{E}$, with $\sigma$ being the optical conductivity, into Eq.~\eqref{eq.P} one gets
\begin{equation}
\sigma(\omega)=4i\omega\sigma_0 \sum_j \frac{{\tilde{\mathbf{d}}^*_j \otimes \tilde{\mathbf{d}}_j}}{\omega-\omega_n+i\gamma/\hbar},
\end{equation}
with $\otimes$ being the outer product, $\sigma_0=e^2/\hbar$, and  $\tilde{\mathbf{d}}_j\equiv \mathbf{d}_j/e$ the dimensionless exciton dipole moment.

\bibliography{ref_new}